\documentclass[12pt]{article}

\usepackage{amsmath,amssymb}
\usepackage[round]{natbib} 
\usepackage{color}
\usepackage{hyperref}
\usepackage{graphicx}

\allowdisplaybreaks
\newtheorem{thm}{Theorem}
\newtheorem{lem}{Lemma}
\newcommand{\bal}[1]{\begin{align} \label{#1} }
\newcommand{\beq}[1]{\begin{equation} \label{#1} }
\newcommand{\eeq}{\end{equation}}

\newcommand{\rf}[1]{(\ref{#1})}

\def\diag{\operatorname{diag}}
\def\det{\operatorname{det}}
\def\tr{\operatorname{tr}}
\def\tens#1{\mathbb{\,#1}}
\def\whbf#1{\widehat{\bf #1}}
\def\ax{\bd p}

\def\bd#1{\mbox{\boldmath$\displaystyle\mathbf{#1}$} }

\begin{document}

\title{ {Euler-Rodrigues and Cayley formulas for  rotation    of  elasticity tensors} }

\author{A. N. Norris\\ \\    Mechanical and Aerospace Engineering, 
	Rutgers University, \\ Piscataway NJ 08854-8058, USA \,\, norris@rutgers.edu \\
	\emph{Dedicated to Professor Michael Hayes on the occasion of his 65th birthday} 
}	\date{}
\maketitle

\begin{abstract} 
It is fairly well known that  rotation in three dimensions can be expressed as a quadratic in a skew symmetric  matrix  via the Euler-Rodrigues formula.  
A generalized Euler-Rodrigues  polynomial of degree $2n$ in a skew symmetric generating matrix is derived for the rotation matrix of tensors of order $n$.  The  Euler-Rodrigues formula for rigid body rotation is recovered by $n=1$.   A Cayley form of the $n^{th}$ order rotation tensor is also derived.  The  representations simplify if there exists some underlying symmetry, as is the case for elasticity tensors such as strain  and the fourth order tensor of elastic moduli.  A new formula is presented for the transformation of elastic moduli under rotation:  as a $21$-vector with a rotation matrix given by  a  polynomial of degree $8$.  Explicit spectral representations are constructed from three vectors: the axis of rotation and two orthogonal bivectors.   The  tensor rotation formulae are related to Cartan decomposition of elastic moduli and  projection onto hexagonal symmetry.

\end{abstract}

\section{Introduction} 


Rigid body rotation about an axis $\ax$, $|\ax | = 1$, is described by the well known Euler-Rodrigues formula  for the rotation matrix 
as a quadratic in the skew symmetric matrix ${\bf P}$,   
\beq{024}
P_{ij} = -\epsilon_{ijk}\, p_k,
\eeq 
 where $\epsilon_{ijk}$ is the third order isotropic alternating tensor\footnote{The summation convention on repeated subscripts is assumed in $\rf{024}_1$ and most elsewhere.}.
Thus, 
\beq{023} {\bf Q}  %
= \exp ( \theta {\bf P}   )
= {\bf I} + \sin\theta\, {\bf P}+ (1 -  \cos\theta) {\bf P}^2,
\eeq
where      $\theta$ is the angle of rotation. 
 Euler  first  derived this formula although  Rodrigues \cite{Rodrigues40} obtained the formula for the composition  of successive finite rotations \cite{Gray80,Cheng89}.   The concise form of the Euler-Rodrigues formula is basically a consequence of the  property 
\beq{2076}
  {\bd P}^3 = -{\bd P}. 
\eeq
Hence, every term in the power series expansion of $ \exp ( \theta {\bf P}   )$ is either    ${\bf P}^{2k}= (-1)^{k+1} {\bf P}^2$ or ${\bf P}^{2k+1} = (-1)^k {\bf P}$, and regrouping  yields \rf{023}.   
Rotation about an axis is  defined  by the  proper orthogonal rigid body rotation matrix ${\bf Q}( \theta ,\ax ) \in SO(3)$,  the  special orthogonal group of matrices satisfying\footnote{${\bf Q}^t$ denotes transpose and ${\bf I}$ is the identity.} 
\beq{131}
{\bf Q} {\bf Q}^t = {\bf Q}^t {\bf Q} = {\bf I}, 
\eeq
 and the special property $\det\,{\bd Q}=1$.   The generating matrix ${\bf P}(\ax ) \in so(3)$ where $so(n)$ denotes the space of skew symmetric matrices or tensors, usually associated with the Lie algebra of the infinitesimal (or Lie) transformation group defined by rotations in $SO(n)$.  Alternatively, $ SO(3)$ is isomorphic to  $so(3)$ via the Cayley form \cite{Bottema79}
\beq{0231} {\bf Q}  
=  ({\bf I} +  {\bf S})\, ({\bf I}-  {\bf S})^{-1} = ({\bf I}-  {\bf S})^{-1}\, ({\bf I} +  {\bf S})
, 
\eeq
where ${\bf S} \in so(3)$ follows from \rf{023} as 
\beq{0232} {\bf S}  %
= \tanh ( \tfrac{\theta}{2} {\bf P}   )
=\tan\! \tfrac{\theta}{2} \, {\bf P}   \, . 
\eeq
The first identity is a simple restatement of the definition in \rf{0231} with ${\bf Q}  
= \exp ( \theta {\bf P})$, while the second is  a more subtle relation that depends upon \rf{2076} and the skew symmetry of ${\bd P}$, and as such is analogous to the Euler-Rodrigues formula. 

   We are concerned  with generalizing these fundamental  formulas for rigid body rotation to the rotation matrices associated with ${\bf Q}$ but which act on second and higher order tensors  in the same way that ${\bf Q}$  transforms a vector as ${\bf v} \rightarrow {\bf v}' = {\bf Q}^t{\bf v}$.  If one takes the view that the rotation transforms a set of orthonormal basis vectors ${\bd e}_1,{\bd e}_2,{\bd e}_3$, then the coordinates of a fixed vector ${\bd v} = v_1{\bd e}_1+v_2{\bd e}_2+v_3{\bd e}_3$ 
relative to this basis  transform  according to $(v_1,v_2,v_3) \rightarrow (v_1',v_2',v_3')$, where $v_i' = Q_{ij}v_j$. 
A tensor $\tens T$ of order (or rank) $n$  has elements, or components relative to a vector basis, 
$T_{ij\ldots kl}$.
Under rotation of the basis vectors by ${\bf Q} $, the elements are transformed by the relations
\beq{591}
T_{ij\ldots kl}' = 
Q_{ij\ldots klpq\ldots rs}\,  
T_{pq\ldots rs}, 
\eeq
where ${\tens Q}= {\tens Q}(  \theta ,{\ax} )$ is a tensor of order $2n$ formed from $n$ combinations of  the underlying rotation,   
\beq{592}
Q_{\underbrace{ij\ldots kl}_{n}\underbrace{pq\ldots rs}_{n}}
\equiv   
\underbrace{Q_{ip} Q_{jq}\ldots Q_{kr} Q_{ls}}_{n}\, . 
\eeq
${\tens Q}$ is also known as the $n$-th Kronecker power of $\bd Q$ \cite{Murnaghan38}. 
Lu and Papadopoulos \cite{Lu98} derived a generalized $n$-th order Euler-Rodrigues formula for ${\tens Q}$ as a polynomial of degree $2n$ in the tensor ${\tens P}$, also of  order $2n$,  defined by  the sum of $n$ terms
\beq{594}
P_{ij\ldots klpq\ldots rs} = 
P_{ip}\delta_{jq}\ldots \delta_{kr}\delta_{ls}
+\delta_{ip}P_{jq}\ldots \delta_{kr}\delta_{ls} + \ldots 
+\delta_{ip}\delta_{jq}\ldots P_{kr}\delta_{ls}
+\delta_{ip}\delta_{jq}\ldots \delta_{kr}P_{ls}\, .  
\eeq
Thus, 
\beq{735}
{\tens Q} = \exp(\theta {\tens P} ) =  P_{2n} (\theta , \, {\tens P}), 
\eeq
where
\beq{736}
P_{2n} (\theta , \, x) = 
\sum\limits_{k=-n}^n\, e^{\lambda_k \theta}
\prod\limits_{\substack{j=-n \\ j\ne k}}^n \bigr( \frac{x - \lambda_j}{\lambda_k-\lambda_j}\bigr), 
\qquad \mbox{with  } \lambda_k =  k\, i ,\quad i = \sqrt{-1}\, .  
\eeq
Lu and Papadopoulos' derivation of \rf{736} is summarized in Section \ref{general}. 
  The case $n=2$ has also been derived in quite different ways by 
Podio-Guidugli  and Virga \cite{pg87} and by Mehrabadi et al. \cite{mcj}. 

The purpose of this paper is  twofold.  First, to  obtain alternative forms of the generalized  Euler-Rodrigues formula and of the associated Cayley form.  Using general properties of  skew symmetric matrices  we will show that the $n$-th order  Euler-Rodrigues formula can be expressed in ways which clearly generalize the classical $n=1$ case.     The difference in approach from that of Lu and Papadopoulos  \cite{Lu98} can be appreciated by noting that their formula \rf{736} follows from the 
fact that the eigenvalues of ${\tens P}$ are $\lambda_k$, $k=-n,  \ldots,  n$, of eq. \rf{736}. 
The present formulation is based on the more fundamental property that ${\tens P}$ has canonical form
\beq{4023} 
{\tens P} = {\tens P}_1+ 2{\tens P}_2+ \ldots + n {\tens P}_n, 
\eeq
where  ${\tens P}_j$ satisfy identities similar to \rf{2076}, and are mutually orthogonal in the sense of tensor product.  
This allows us to express $\exp (\theta {\tens P})$ and $\tanh (\tfrac{\theta}{2} {\tens P})$ in  forms that more clearly generalize \rf{023} and \rf{0232}.  Furthermore, the base tensors ${\tens P}_j$ can be expressed in terms of powers of ${\tens P}$, which imply explicit equations for $\tens Q$ and the related $\tens S$.    We will derive these equations in Section \ref{general}.  

The second objective  is to derive related  results for  rotation of tensors in elasticity.  The underlying symmetry of the physical tensors, strain, elastic moduli, etc., allow further simplification from the $2n-$ tensors ${\tens Q}\in SO(3^n)$  and  ${\tens P}\in so(3^n)$, where the nominal dimensionality $3^n$ assumes no symmetry.   Thus, Mehrabadi et al. \cite{mcj} expressed the rotation tensor  associated with strain, $n=2$, in term of second order tensors in 6-dimensions, significantly reducing the number of components required.  By analogy, 
the rotation matrix for fourth order elastic stiffness tensors  is derived here as a second order tensor in 21-dimensions
$\widetilde{\bf Q}( \theta ,\ax  ) = \exp \big(\theta \widetilde{P}\big)\in SO(21)$ where $\widetilde{P}\in so(21)$.   This provides an alternative to existing methods for calculating elastic moduli under a change of basis, e.g. using  Bond transformation matrices \cite{AuldI} or other methods based on   representations of the moduli as elements of $6\times 6$ symmetric matrices \cite{mcj}.  Viewing the elastic moduli as a $21$-vector is the simplest approach for some purposes, such as projection onto particular symmetries \cite{Helbig95,Browaeys04}.   In fact, we will see that the generalized Euler-Rodrigues formula leads to a natural method for projecting onto hexagonal symmetry defined by the axis ${\bd p}$, as an explicit expression for the projector appears quite naturally. 

The  third and final result is a general formulation of the so-called Cartan decomposition \cite{FV} of tensor rotation.  The action of  $\bd Q$ of \rf{023} on a vector leaves the component parallel to the axis $\bd p$ unchanged, and the component perpendicular to  it rotates through angle $\theta$.  The latter decreases the part perpendicular to  $\bd p$ to $\cos\theta$ of its original value, and introduces $\sin\theta$ times the same magnitude but in in a direction orthogonal to both the vector and the axis.  This simple geometrical interpretation   generalizes for tensors, in that we can identify ``components" that remain unchanged, and others that rotate according to   $\cos j\theta$, $\sin j\theta$, $j=1,2,\ldots  , n$.  The components form subspaces that rotate independently of one another, just as the axial and perpendicular components of the vector do for $n=1$.  This is the essence of  Cartan decomposition of tensors, see \cite{FV} for further details.  Our main result is that the components and the subspaces can be easily identified and defined using the properties of $\tens P$ and $\tens Q$ that follow from \rf{4023}.

We begin with a summary of the main findings in Section \ref{main}, with general results applicable to tensors of arbitrary order described in Section \ref{general}.  Further details of the proofs are given in Section \ref{eigs}.  Applications to elasticity are described  in Section \ref{apps} where the 21 dimensional rotation of elastic moduli is derived.  Finally, the relation between the generalized Euler-Rodrigues formula and the Cartan decomposition of elasticity tensors is discussed in  Section \ref{Discussion}.  

A note on notation: tensors of order $n$ are denoted $\tens P$, $\tens Q$, etc.  
Vectors and matrices in 3D are normally lower and capital boldface, e.g. $\bd p$, $\bd Q$, while quantities in $6$-dimensions are denoted e.g. $\whbf{Q}$, and in $21$-dimensions  as $\widetilde{Q}$. 

\section{Summary of results}\label{main}

Our principal result is   
\begin{thm}\label{thm1}
The rotation tensor ${\tens Q} ( \theta ,\ax )= \exp\big(\theta {\tens P} \big) $ 
has the form 
\beq{344}
{\tens Q}  = {\tens I}
 + 
\sum\limits_{k=1}^n\,  \big( \sin k\theta \,{\tens P}_k  + 
 (1- \cos  k\theta )   {\tens P}_k^2 \big) 
\eeq
where ${\tens P}_k$, $k=1,2,\ldots , n$ are mutually orthogonal skew-symmetric tensors that partition ${\tens P}$ and have the same basic property as $\bd P$ in \rf{2076}, 
\begin{subequations}\label{106}
\bal{106a} 
&{\tens P} = {\tens P}_1+ 2{\tens P}_2+ \ldots + n {\tens P}_n, 
\\
&{\tens P}_i{\tens P}_j = {\tens P}_j{\tens P}_i = 0,\quad i\ne j , 
\label{106b}
\\
&{\tens P}_i^3 = -{\tens P}_i\, .  
\label{106c}
\end{align}
\end{subequations}
The Cayley form for $\tens Q$ is
\beq{0233} {\tens Q}  
=  ({\tens I} +  {\tens S})\, ({\tens I}-  {\tens S})^{-1} = ({\tens I}-  {\tens S})^{-1}\, ({\tens I} +  {\tens S})
, 
\eeq
where 
\beq{0234} {\tens S}  
= \sum\limits_{k=1}^n\, \tan (k\tfrac{\theta}{2}) \, {\tens P}_k   \, . 
\eeq
The tensors ${\tens P}_k$ are uniquely defined by polynomials of  ${\tens P}$ of degree $2n-1$, 
\beq{345}
{\tens P}_k = p_{n,k}({\tens P}), 
\qquad
\mbox{where}\quad 
p_{n,k}(x) = \frac{x}{k} \, 
\prod\limits_{\substack{j=1 \\ j\ne k}}^n \bigr( \frac{x^2 +j^2}{j^2-k^2}\bigr),\quad 1\le k\le n.
\eeq
The polynomial representation  ${\tens Q} =P_{2n} (\theta , {\tens P})$ 
has several alternative forms, three of which   are  as follows: 
\begin{subequations}\label{5191}
\bal{5191a}
P_{2n} (\theta , \, x) &=  1  +   
\sum\limits_{k=1}^n\,  \frac{2(-1)^{k+1}} {(n-k)!\,(n+k)! }  \, 
\big[ k \sin  k\theta  \,x + ( 1- \cos k\theta ) \,x^2  \big] 
\, 
 { \prod\limits_{\substack{j=1 \\ j\ne k}}^n (  x^2 + j^2) } 
 \\
 &= 1 
 + \sum\limits_{k=1}^n\, \frac{ (2\sin\frac{\theta}{2})^{2k-1} }{k(2k-1)!}\,
 \big( k\cos\tfrac{\theta}{2} \, x + \sin\tfrac{\theta}{2} x^2)
 \,  \prod\limits_{l=1}^{k-1} (x^2 + l^2)  
 \label{5191b}
 \\
 &= 1 +\sin\theta\, x \bigr( 
1 + \frac{X_1}{1}\bigr(
1 + \frac{X_2}{2}\big(
1 + \frac{X_3}{3}(1 + \ldots \frac{X_{n-1}}{n-1}
)\big)\bigr)\bigr)
 \nonumber \\ & \qquad 
 + (1 -  \cos\theta)\, x^2 \bigr(
1 + \frac{X_1}{2}\bigr(
1 + \frac{X_2}{3}\big(
1 + \frac{X_3}{4}(1 + \ldots \frac{X_{n-1}}{n}
)\big)\bigr)\bigr)
 \, , 
  \label{5191c}
\end{align}
\end{subequations}
where 
\beq{711}
X_j = (1-\cos \theta) (x^2+j^2)/({2j+1}) .
\eeq
\end{thm}

We note that
 $n=1,2,3,4$  corresponds to rotation of vectors, second, third  and fourth order tensors, respectively.  All these are covered by considering $n=4$ in Theorem \ref{thm1}, for which the series \rf{5191c} and \rf{5191a} are, respectively, 
 \bal{433} 
P_{8} (\theta ,  x) 
= &1 + \sin\theta\, x\bigg( 
1 + \frac{ (1 -  \cos\theta)}{1.3} (x^2 + 1)\bigg[
1 + \frac{ (1 -  \cos\theta)}{2.5} (x^2 + 4)\bigr(
1 + \frac{ (1 -  \cos\theta)}{3.7} (x^2 + 9)\bigr)\bigg]\bigg)
 \nonumber \\ 
 &
 + (1 -  \cos\theta)\, x^2 \bigg(
 1 + \frac{ (1 -  \cos\theta)}{2.3} (x^2 + 1)\bigg[
 1 + \frac{ (1 -  \cos\theta)}{3.5} (x^2 + 4)\bigr( 
1 + \frac{ (1 -  \cos\theta)}{4.7} (x^2 + 9)\bigr)\bigg]\bigg)\,
\nonumber \\
=& \, 1 + 
\frac1{5!3} \big[ \sin\theta\, x+ (1 -  \cos\theta) x^2\big] 
(x^2 +4) (x^2 +9) (x^2 +16) 
\nonumber \\
&  \quad - \frac{1}{6!}\big[ 2\sin 2\theta\, x+ (1 -  \cos 2\theta) x^2\big] (x^2 +1) (x^2 +9) (x^2 +16) 
\nonumber \\
&  \quad + \frac{2}{7!}\big[ 3\sin 3\theta\, x+ (1 -  \cos 3\theta) x^2\big] (x^2 +1) (x^2 +4) (x^2 +16) 
\nonumber \\
& \quad  - \frac{2}{8!}\big[ 4\sin 4\theta\, x+ (1 -  \cos 4\theta) x^2\big] (x^2 +1) (x^2 +4) (x^2 +9) \,   . 
\end{align}
This expression includes all the Euler-Rodrigues type formulas for tensors of order $n\le 4$, on account of the 
characteristic polynomial equation for ${\tens P}$  of degree $2n+1$, 
\beq{13}
 {\tens P}({\tens P}^2+{\tens I})({\tens P}^2+4{\tens I})\ldots ({\tens P}^2+n^2{\tens I})= 0\, . 
\eeq 
Thus, the last line in \rf{433} vanishes for $n=3$, the last two for $n=2$, and all but the first line for $n=1$,  while the terms that remain can be simplified using \rf{13}. 
The formulas for $P_{2n}$, $n=2$ and $3$, are 
\begin{subequations}\label{14}
\bal{14a}
P_{4}(\theta, x) = &1 + \sin\theta\, x 
\bigr( 1 + \frac13 (1 -  \cos\theta) (x^2 + 1)\bigr) 
+ (1 -  \cos\theta)\, x^2 \bigr( 1 + \frac16(1 -  \cos\theta) (x^2 + 1)\bigr) ,
\\
P_{6}(\theta, x)= &1 + \sin\theta\, x\bigr( 
1 + \frac13{ (1 -  \cos\theta)} (x^2 + 1)\bigr[
1 + \frac1{10}{ (1 -  \cos\theta)} (x^2 + 4)\bigr]\bigr)
 \nonumber \\  &
 + (1 -  \cos\theta)\, x^2 \bigr(
 1 + \frac16 { (1 -  \cos\theta)}  (x^2 + 1)\bigr[
 1 + \frac1{15} { (1 -  \cos\theta)} (x^2 + 4)\bigr]\bigr) \, . 
\end{align}
\end{subequations}
 These identities  apply for $n\le2$ and $n\le3$, respectively, since they reduce to, e.g. the $n=1$ formula using \rf{13}. 
 
The second main result is a matrix representation for simplified forms of  $\tens P$ and $\tens Q$ for 4th order elasticity tensors.  Although of fourth order,  symmetries reduce the maximum number of elements from $3^4$ to $21$, and hence the moduli can be described by a $21$-vector and $\tens P$ and $\tens Q$ by $21\times 21$ matrices.  This reduction in matrix size is similar to the $6\times 6$ representation of Mehrabadi et al. \cite{mcj} for symmetric second order tensors.  The case of third order tensors ($n=3$) is not discussed here to the same degree of detail as for second and fourth order, but it could  be considered in the same manner. 
The following Theorem provides the matrices for $n=1,2,4$, 
corresponding to rotation of vectors, symmetric second order tensors, and fourth order elasticity   tensors, respectively.   

 \begin{thm}\label{thm2}
 The skew-symmetric matrices and the associated rotation matrices defined by the tensors $({\tens P}{,\tens Q})$  for $n=1,2,4$, are given by  $({\bf P},{\bf Q})$, $(\whbf{\bf P},\whbf{\bf Q})$ and $(\widetilde{\bf P},\widetilde{\bf Q})$, respectively. 
The $m\times m$, $m=3,6,21$,  skew symmetric generator matrix   for  each case has the form 
\beq{21345}
P = R-R^t,
\eeq
where 
\begin{subequations}\label{2133}
\begin{align}
\qquad {\bf R} &=  - {\bf X} , \qquad  \mbox{for vectors},\ n=1, &
\\ && \nonumber \\
\whbf{R}  &=  \begin{pmatrix}
{\bf 0} & \sqrt{2}{\bf Y} 
       \\ 
\sqrt{2}{\bf Z} & {\bf X}  
\end{pmatrix} , \qquad \mbox{for symmetric tensors}, \  n=2, &
\\ && \nonumber \\
\widetilde{\bf R} & =  \begin{pmatrix} 
{\bf 0} & {\bf 0} & {\bf 0} & {\bf 0} &2{\bf Y}  & {\bf 0} &{\bf 0} 
					\\ 
{\bf 0} & {\bf 0} & {\bf 0} & -\sqrt{2}{\bf Y}  & {\bf 0}  & \sqrt{2}{\bf N}  &{\bf 0} 
					\\ 
{\bf 0} & {\bf 0} & {\bf 0} & {\bf 0} & {\bf 0} & 2{\bf N}  & -\sqrt{2}{\bf Y}  
					\\ 
{\bf 0} & -\sqrt{2}{\bf Z} & {\bf 0} & {\bf 0} & {\bf X} &{\bf 0} &  -\sqrt{2}{\bf X}  
					\\ 
{\bf 0}  &\sqrt{2}{\bf N} & 2{\bf N} & {\bf 0} & {\bf 0} & {\bf X}  & {\bf 0}
					\\ 
 2{\bf Z}   & {\bf 0} & {\bf 0} &   {\bf X} & {\bf 0} & {\bf 0} &   \sqrt{2}{\bf X}
   				\\ 
{\bf 0} &{\bf 0} & -\sqrt{2}{\bf Z}  & - \sqrt{2}{\bf X}   &\sqrt{2}{\bf X} & {\bf 0} &  - {\bf X}
\end{pmatrix}
,  \   
\begin{split}  & \mbox{for         elasticity} \\ & \mbox{tensors}, \ n=4, 
\end{split} &
\qquad \qquad
\end{align}
\end{subequations}
with ${\bf 0} = {\bf 0}_{3\times 3}$ and 
\beq{509}
{\bf X} = \begin{pmatrix}
0  & p_3  & 0 \\ 
0 & 0& p_1 \\ 
 p_2 & 0 & 0
\end{pmatrix} ,
\quad
 {\bf Y} = \begin{pmatrix} 
0  & p_2  & 0 \\ 
0 & 0& p_3 \\ 
 p_1 & 0 & 0
\end{pmatrix} ,
\quad
 {\bf Z} = \begin{pmatrix} 
0  & p_1  & 0 \\ 
0 & 0& p_2 \\ 
 p_3 & 0 & 0 
\end{pmatrix} ,
\quad
 {\bf N} = \begin{pmatrix} 
p_1  & 0  & 0 \\ 
0 & p_2& 0 \\ 
0 & 0 & p_3
\end{pmatrix} \, . 
\eeq
In each case, the $Q$-matrix is given by Theorem \ref{thm1}, and the tensors  acted on by $Q$ are $m$-vectors  which transform like $v'= Qv$. 
\end{thm}

The specific form of the $6$- and $21$-vectors are given in Section \ref{apps}. 

The third and final  result identifies subspaces that are closed under rotation:  
\begin{thm}\label{thm3}
The Cartan components of  an $n$-th order tensor $\tens T$ are defined as
\beq{644}
{\tens T}_0 =  {\tens M}_0{\tens T},\qquad
{\tens T}_j   \equiv {\tens M}_j{\tens T} ,
\quad
{\tens R}_j   \equiv {\tens P}_j{\tens T} ,
\quad j=1,2,\ldots, n. 
\eeq
where the   $n\negthinspace +\negthinspace 1$ 
 symmetric projection tensors, ${\tens M}_k$, $k=0, 1, \ldots, n$,  are    
\beq{640}
{\tens M}_0 \equiv {\tens I} + {\tens P}_1^2 + {\tens P}_2^2 + \ldots + {\tens P}_n^2 ,
\qquad {\tens M}_i = -{\tens P}_i^2, \quad i=1,2,\ldots, n. 
\eeq
They satisfy, for $ 0\le k\le n$ and $1\le i \le n$, 
\begin{subequations}\label{641}
\bal{641a} 
&{\tens M}_i{\tens M}_k ={\tens M}_k{\tens M}_i = {\tens M}_k{\tens P}_i = {\tens P}_i{\tens M}_k = 0,\quad i\ne k , 
\\
&{\tens M}_k^2 = {\tens M}_k, 
\qquad {\tens M}_i{\tens P}_i = {\tens P}_i{\tens M}_i = {\tens P}_i , 
\label{641b}
\\
&{\tens M}_k = m_{n,k} \big( {\tens P}\big),\qquad 
m_{n,k}(x) = \frac{(-1)^{k}\, (2-\delta_{k0}) }{(n-k)!(n+k)! } \, 
\prod\limits_{\substack{j=0 \\ j\ne k}}^n ( x^2 +j^2) . 
\label{641c}
\end{align}
\end{subequations}

The projection ${\tens T}_0$ along with 
the   $n-$pairs $\{ {\tens T}_j,  \, {\tens R}_j\} $, $j=1,2,\ldots, n$ define $n+1$ 
subspaces  that  are closed under rotation:     
\beq{082}
 { \tens Q} { \tens T}_0 =  { \tens T}_0,
\qquad
 { \tens Q} { \tens T}_j = \cos j\theta  { \tens T}_j + \sin j\theta  { \tens R}_j,
\quad
 { \tens Q} { \tens R}_j = \cos j\theta  { \tens R}_j - \sin j\theta  { \tens T}_j,
\quad j=1,2,\ldots , n. 
 \eeq
\end{thm}
The meaning will become more apparent by example, as we consider the various cases of tensors of order $n=1, 2, $ and $4$ in Section \ref{Discussion}. 

\section{General theory for tensors}\label{general}

\subsection{$P$'s  and $Q$'s}

The transpose of a $2n$-th order tensor is defined by interchanging the first and last $n$ indices.  In particular,  the skew symmetry of ${\bd P}$ and the definition of ${\tens P}$ in \rf{594} implies that 
 ${\tens P}^t = -{\tens P}$.  Based on the definitions of $\tens Q$ in \rf{592} and the properties of the fundamental rotation ${\bd Q}\in SO(3)$, $\tens Q$ of \rf{735} satisfies
\beq{598}
 \frac{{\rm d} {\tens Q}}{{\rm d} \theta} = {\tens P}{\tens Q} , 
\eeq
where the product of two tensors of  order $2n$  is another tensor of  order $2n$ defined by contracting over $n$ indices, 
$(AB)_{ij\ldots klpq\ldots rs }  = A_{ij\ldots klab\ldots cd } \, B_{ab\ldots cdpq\ldots rs }$.  
This gives meaning to the representation 
\beq{599} 
{\tens Q} = e^{\theta {\tens P}}.  
\eeq
Based on the skew symmetry of ${\tens P}$  it follows that 
\beq{600}
{\tens Q}{\tens Q}^t ={\tens Q}^t{\tens Q} ={\tens I},  \quad \mbox{where  }I_{ij\ldots klpq\ldots rs} 
= \overbrace{\delta_{ip} \delta_{jq}\ldots \delta_{kr} \delta_{ls}}^{n}\, . 
\eeq
The isomorphism between the space of $n$-th order tensors and a vector space of dimension $3^n$ implies  that  ${\tens Q}\in SO(3^n)$ and ${\tens P}\in so(3^n)$. 

The derivation of the main result in Theorem \ref{thm1} is   outlined next using a series of Lemmas.  Details of the proof are provided below and in the Appendix. 

\begin{lem}\label{lem1}
For any $N\ge 1$, a given  non-zero $N\times N$ skew symmetric matrix $B$ has $2m\le N$ distinct non-zero eigenvalues of the form $\{\pm i c_1, \, \pm i c_2,\ldots , \pm i c_m\}$,   $c_1,\ldots, c_m$  real, and  $B$ has the representation  
\begin{subequations}\label{05}
\bal{05a} 
&B = c_1B_1+ c_2B_2+ \ldots + c_m B_m, 
\\
&B_iB_j = B_jB_i = 0,\quad i\ne j ,
\\
&B_i^3 = -B_i. 
\end{align}
\end{subequations}
\end{lem}
Lemma \ref{lem1} is essentially  Theorem 2.2 of Gallier and Xu \cite{Gallier02}, who also noted the immediate  corollary 
\bal{0502}
e^{B} &= I + \sum\limits_{k=1}^\infty \frac1{k!}\big(c_1B_1+ c_2B_2+ \ldots + c_m B_m\big)^k
\nonumber \\
&= I + \sum\limits_{k=1}^\infty \frac1{k!}\big(c_1^kB_1^k+ c_2^kB_2^k+ \ldots + c_m^k B_m^k\big) 
\nonumber \\
&= e^{c_1B_1}+e^{c_2B_2}+\ldots +e^{c_m B_m} -(m-1)I
\nonumber \\
&= I  + 
\sum\limits_{j=1}^m\,  \big( \sin c_j B_j + 
 (1- \cos c_j )   B_j^2 \big) \, . 
\end{align}
Thus, the exponential of any skew symmetric matrix has the form of a sum of Euler-Rodrigues terms. 
\begin{lem}\label{lem2} 
The elements of the decomposition of Lemma 1 can be expressed in terms of the matrix $B$ and its eigenvalues by  
\beq{503}
B_j = \frac{ B }{ c_j} 
\prod\limits_{\substack{k=1 \\ k\ne j}}^m
\bigr(\frac{ B^2 + c_k^2 }{c_k^2 - c_j^2}\bigr)   \, ,  
\eeq
\end{lem}
 Equation \rf{503} may be obtained by starting with 
\beq{5034}
B(B^2+c_k^2) = \sum\limits_{\substack{j=1 \\ j\ne k}}^m\, (c_k^2 - c_j^2)c_j B_j , 
\eeq
and iterating until a single $B_j$ remains.  Equation \rf{503} in combination with \rf{0502} and  $B_j^2 = c_j^{-1}BB_j$ implies 
that $e^{B} $ can be expressed as a polynomial of degree $2m$ in $B$, 
\beq{05030}
e^{B} 
= I  + 
\sum\limits_{j=1}^m\,  
\big[ c_j \sin  c_j  \,B + ( 1- \cos  c_j ) \,B^2  \big] 
\, c_j^{-2} \, 
  \prod\limits_{\substack{k=1 \\ k\ne j}}^m 
 \bigr(\frac{ B^2 + c_k^2 }{c_k^2 - c_j^2}\bigr) \, . 
\eeq
   We are now ready to consider  $\exp( \theta {\tens P} )$.
\begin{lem}\label{lem3} 
The non-zero eigenvalues of the skew symmetric tensor $\tens P$  defined in eq. \rf{594} are
\[ \{ i,\, -i, \,  2i, \,  -2i, \, \ldots , ni, - ni\}. 
\]
\end{lem}
This follows from Zheng and Spence \cite{Zheng93} or Lu and Papadopoulos \cite{Lu98}, and is discussed in detail in Section \ref{eigs}. 
${\tens P}$ therefore satisfies the characteristic equation \rf{13}, and  has the properties 
\begin{subequations}\label{06}
\bal{06a} 
&{\tens P} = {\tens P}_1+ 2{\tens P}_2+ \ldots + n {\tens P}_n, 
\\
&{\tens P}_i{\tens P}_j = {\tens P}_j{\tens P}_i = 0,\quad i\ne j , 
\label{06b}
\\
&{\tens P}_i^3 = -{\tens P}_i\, .  
\label{06c}
\end{align}
\end{subequations}
Equation \rf{06a} is the canonical decomposition of ${\tens P}$ into orthogonal components each of which has properties like the fundamental $\bd P$ in \rf{2076}, although the subspace associated with ${\tens P}_i$ can be  multidimensional.  The polynomials $p_{n,k}(x)$ of eq. \rf{345}  follow from \rf{503}, or alternatively, 
\beq{3450}
p_{n,k}(x) = \frac{(-1)^{k+1}\, 2k x }{(n-k)!(n+k)! } \, 
\prod\limits_{\substack{j=1 \\ j\ne k}}^n ( x^2 +j^2),\quad 1\le k\le n.
\eeq 
We will examine the particular form of the ${\tens P}_i$ for elasticity tensors of order $n=2$ and $n=4$ in Section \ref{apps}.  For now we note the consequence of Lemmas \ref{lem2} and \ref{lem3}, 
\beq{0501}
e^{\theta {\tens P}} 
=  {\tens I}  + 
\sum\limits_{k=1}^n\,  
\big[ k \sin  k\theta  \,{\tens P} + ( 1- \cos k\theta ) \,{\tens P}^2  \big] 
\, k^{-2} \, 
  \prod\limits_{\substack{j=1 \\ j\ne k}}^n \bigr( \frac{ {\tens P}^2 + j^2 {\tens I} } {j^2 - k^2} \bigr)\, . 
\eeq
Together with  \rf{599}  this implies  the first expression \rf{5191a}  in Theorem \ref{thm1}.  The alternative identities \rf{5191b} and \rf{5191c} are  derived in the Appendix. 

\begin{lem}\label{lem4} 
${\tens Q}$ may be expressed in Cayley form
\beq{0235} {\tens Q}  
=  ({\tens I} +  {\tens S})\, ({\tens I}-  {\tens S})^{-1} = ({\tens I}-  {\tens S})^{-1}\, ({\tens I} +  {\tens S})
, 
\eeq
where  the $2n$-th order  skew symmetric  tensor ${\tens S}$ is 
\beq{0236} {\tens S}  
= \tanh ( \tfrac{\theta}{2} {\tens P}   ) 
= \sum\limits_{k=1}^n\, \tan (k\tfrac{\theta}{2}) \, {\tens P}_k   \, . 
\eeq
\end{lem}
This follows by inverting \rf{0235} and using the expression \rf{344} for $\tens Q$, 
\beq{0237} {\tens S}  
=  {\tens I} - 2({\tens I} +  {\tens Q})^{-1}
= {\tens I} - \big[ {\tens I} + 
\sum\limits_{k=1}^n\,  \frac12 \big(  \sin k\theta \,{\tens P}_k  + 
 (1- \cos  k\theta )   {\tens P}_k^2 \big)\big]^{-1}
, 
\eeq
and then applying the following identity, 
\beq{0238} 
\big[ {\tens I} + 
\sum\limits_{k=1}^n\,  \big(  a_k {\tens P}_k  + 
 (1- b_k )   {\tens P}_k^2 \big)\big]^{-1}
= {\tens I} + 
\sum\limits_{k=1}^n\,  \big(  -\frac{ a_k}{a_k^2 + b_k^2} {\tens P}_k  + 
 (1- \frac{b_k}{a_k^2 + b_k^2} )   {\tens P}_k^2 \big)\, . 
\eeq
Note that Lemma \ref{lem4} combined with the obvious result $\rf{0236}_1$ implies that, for arbitrary $\phi$, 
\beq{076}
\tanh (\phi {\tens P} ) = 
 \sum\limits_{k=1}^n\, \tan (k\phi) \, {\tens P}_k .
\eeq
This is a far more general statement than the simple partition of \rf{4023}, and in fact may be shown to be equivalent to eq. \rf{06}. 

\subsection{Equivalence with the Lu and Papadopoulos formula}
We now show that \rf{0501} agrees with the polynomial of Lu and Papadopoulos \cite{Lu98}.  They derived  eq. \rf{736} directly using Sylvester's interpolation formula (p. 437 of Horn and Johnson \cite{Horn91}). 
Thus, the function $f({\bd A})$ defined by a power series for a  matrix ${\bd A}$ that is diagonalizable can be expressed in terms of its distinct eigenvalues $\lambda_1,\lambda_2,\ldots \lambda_r$  as  $f({\bd A}) = p({\bd A})$, 
where $p(x)$ is  
\beq{5271}
p(x) = \sum\limits_{i=1}^r\, f(\lambda_i)\, L_i(x) ,
\qquad
L_i(x) = \prod\limits_{\substack{j=1 \\ j\ne i}}^r\, 
\bigr( \frac{x-\lambda_j}{\lambda_i-\lambda_j}\big)\, . 
\eeq
$L_i(x)$ are the Legendre interpolation polynomials  and $p(x)$ is the unique polynomial of degree $r-1$ with the property $p(\lambda_i)= f(\lambda_i)$. 
Equation \rf{736} follows from the  explicit form of the spectrum of $\tens P$. 
The right member of \rf{736} 
can be rewritten  in the following form by combining the terms $e^{\pm ik\theta}$, 
\beq{62}
P_{2n} (\theta , \, x)  =   \frac1{(n!)^2} \prod\limits_{j=1 }^n (  x^2 + j^2) 
  +   
\sum\limits_{k=1}^n\,  \frac{2(-1)^{k+1}} {(n-k)!\,(n+k)! }  \, 
\bigr( k \sin  k\theta  \,x - \cos k\theta  \,x^2  \bigr) 
\,   \prod\limits_{\substack{j=1 \\ j\ne k}}^n (  x^2 + j^2)  \, . 
\eeq
The first term on the right hand side  can be written in a form which agrees with \rf{5191a} by considering the partial fraction expansion  of $1/\Lambda $ where 
\beq{6300}
\Lambda (x)= \prod\limits_{k=1 }^n ( x^2 + k^2  )\, .   
\eeq
Thus, as  may be checked by comparing residues, for example, 
\bal{6302}
\frac1{\Lambda }
&=  
\sum\limits_{k=1}^n\,  \frac{2 (-1)^{k+1}} {(n-k)!\,(n+k)! }  \frac{k^2}{  x^2 + k^2}  
\nonumber \\    
&= \sum\limits_{k=1}^n\,  \frac{2  (-1)^{k+1}} {(n-k)!\,(n+k)! } 
 - 
\sum\limits_{k=1}^n\,  \frac{2 (-1)^{k+1}} {(n-k)!\,(n+k)! }  \frac{x^2}{  x^2 + k^2}  \, .   
\end{align}
The first term in the right member is $1/\Lambda (0) = 1/(n!)^2$.  Multiplying by 
$\Lambda (x)/\Lambda (0)$ implies the identity 
\beq{63}
 \frac1{(n!)^2} \prod\limits_{j=1 }^n (  x^2 + j^2) 
 = 1 + 
\sum\limits_{k=1}^n\,  \frac{2x^2 (-1)^{k+1}} {(n-k)!\,(n+k)! }  \, 
  \prod\limits_{\substack{j=1 \\ j\ne k}}^n (  x^2 + j^2)  \, , 
\eeq
which combined with \rf{62} allows us  recover the form \rf{5191a}.   This  transformation from the interpolating polynomial \rf{736} to the alternative forms in eq. \rf{5191} is rigorous but it  does not capture the physical basis of the latter.  We will find the identity \rf{63} useful in Section \ref{Discussion}. 

\section{Eigenvalues of $\tens P$ and application to second order tensors}\label{eigs}

The key quantity in the polynomial representation of $\tens Q$ is the set  of distinct non-zero eigenvalues of $\tens P$, Lemma \ref{lem3}. We prove this by construction, starting with the case of $n=1$. 

\subsection{Rotation of vectors, $n=1$}
 
Let $\{ \ax, {\bf q}, {\bf r}\}$ form an orthonormal triad of vectors, then the eigenvalues and eigenvectors of $\bf P$ are as follows
\beq{ev3}
\begin{matrix}
\mbox{eigenvalue}  & \mbox{eigenvector}      
\\ 
0 & \ax
\\ 
\pm i & \qquad {\bf v}_{\pm} \equiv \frac1{\sqrt{2}}(i {\bf q} \pm {\bf r} ) \qquad
\end{matrix} 
\eeq
These may be checked  using the  properties of the third order alternating tensor.  Thus, $P_{ij}q_j = r_i$,  $P_{ij}r_j = -q_i$, implying ${\bf P}{\bf v}_\pm = \pm i {\bf v}_\pm$. 
Hence,  the spectral representation of ${\bf P}$ is 
\beq{p11}
 {\bf P} = i  {\bf v}_+{\bf v}_+^* - i  {\bf v}_-{\bf v}_-^*\, , 
 \eeq
 where $^*$ denotes complex conjugate.  It also denotes transpose if we view eq. \rf{p11} in vector/matrix format. We can also think of   this as a tensorial representation  in which   terms such as ${\bf v}_-{\bf v}_-^*$ stand for the Hermitian dyadic ${\bf v}_-\otimes {\bf v}_-^*$, however, for simplicity of notation we do not use   dyadic notation further but take the view that dyadics are obvious from the context. 
 
 The  bivectors \cite{bhbiv} ${\bf v}_{\pm}$, together with the axis $\bd p$, will serve as the building blocks for spectral representations of $\tens P$ for $n>1$. 
 
  Referring to eq. \rf{06}, we see that ${\tens P}={\tens P}_1$ in this case. 
 Equation \rf{p11} allows us to evaluate powers and other functions of ${\bf P}$.  Thus, in turn, 
 \begin{subequations}\label{p117}
 \bal{p117a}
 {\bf P}^2 &= -  {\bf v}_+{\bf v}_+^* -   {\bf v}_-{\bf v}_-^* \, , 
 \\ 
 {\bf P}^2 +{\bf I}  &=   \ax\ax,
 \\ 
 {\bf P}({\bf P}^2 +{\bf I})  &=  0, 
 \end{align}
 \end{subequations}
from which  the characteristic equation \rf{2076} follows. 

\subsection{Rotation of second order tensors, $n=2$}

In this case the tensors $\tens P$ and $\tens Q$ are fourth order with, see 
 eq. \rf{594}, 
\beq{5941}
P_{ijkl} = 
P_{ik}\delta_{jl}+ \delta_{ik}P_{jl}\, .  
\eeq
Equation \rf{345} implies 
\beq{00821}
 {\tens P}_1 = {\tens P} ({\tens P}^2+4)/3, 
 \qquad
 {\tens P}_2= -{\tens P} ({\tens P}^2+1)/6,  
\eeq
which clearly satisfy the decomposition  \rf{06a}
\beq{0081}
 {\tens P}={\tens P}_1+2{\tens P}_2.  
 \eeq
The skew symmetric tensor $\tens S$ of \rf{0234} may be expressed 
\beq{820}
 {\tens S} = \big( \tfrac43 \tan \tfrac{\theta}{2} -\tfrac16 \tan \tfrac{3\theta}{2}\big) 
 {\tens P} 
  + 
 \big( \tfrac13 \tan \tfrac{\theta}{2} -\tfrac16 \tan \tfrac{3\theta}{2}\big) 
 {\tens P}^3.  
\eeq

Consider the product of the second order tensor (dyad) $\ax {\bf v}_\pm$ with $\tens P$.  Thus,  
${\tens P}\ax {\bf v}_\pm = \pm i \ax {\bf v}_\pm $ which together with ${\tens P}{\bf v}_\pm \ax = \pm i {\bf v}_\pm \ax $  yields 4 eigenvectors, two pairs with eigenvalue $i$ and $-i$.  The dyadics ${\bf v}_\pm{\bf v}_\pm$ are also eigenvectors and have eigenvalues $\pm 2i$. The remaining 3 eigenvectors of $\tens P$ are null vectors  which  can be identified as $\ax \ax$, ${\bf v}_+{\bf v}_-$ and ${\bf v}_-{\bf v}_+$.  This completes the $3^n=9$ eigenvalues and eigenvectors, and shows that the non-zero eigenvalues are $\{i(2),-i(2),2i,-2i\}$ where the number in parenthesis indicates the multiplicity.  The tensors ${\tens P}_1$ and ${\tens P}_2$ can be expressed explicitly in terms of the eigenvectors, 
 \beq{0082}
 {\tens P}_1 = i \ax {\bf v}_+\ax {\bf v}_+^* +i  {\bf v}_+\ax{\bf v}_+^*\ax 
 			-i \ax {\bf v}_-\ax {\bf v}_-^* -i  {\bf v}_-\ax{\bf v}_-^*\ax,
 \qquad
 {\tens P}_2= i {\bf v}_+{\bf v}_+{\bf v}_+^*{\bf v}_+^*-i {\bf v}_-{\bf v}_-{\bf v}_-^*{\bf v}_-^*. 
\eeq
It is straightforward to show that these satisfy the  conditions \rf{06}.  

We note in passing that zero eigenvalues correspond to tensors of order $n$ with rotational or transversely isotropic symmetry about the ${\ax}$ axis.  This demonstrates that there are three second order basis tensors for  transversely isotropic symmetry.  We will discuss this aspect in greater detail in Section \ref{Discussion} in the context of elasticity tensors. 

Both ${\tens P}$ and ${\tens Q}$ correspond to $3^n\times 3^n$ or $9\times 9$ matrices, which we denote $P$ and $Q$, with the precise form dependent on how we choose to represent second order tensors as $9$-vectors.  To be specific, let $T_{ij}$ be the components of a second order tensor (not symmetric in general), and define the $9$-vector
\beq{036}
T = \big( T_{11},\, T_{22},\, T_{33},\, T_{23},\, T_{31},\, T_{12},\, T_{32},\, T_{13},\, T_{21}
\big)^t\, . 
\eeq
The form of $P$ follows from the expansion \rf{591} for small $\theta$, or from \rf{594}, as
\beq{037}
P   =  \begin{pmatrix} 
0 & 0 & 0 & 	        0 & p_2 & -p_3  &     0 & p_2 & -p_3
\\ 
0 & 0 & 0 & 	        -p_1 & 0 & p_3  &     -p_1 & 0 & p_3
\\ 
0 & 0 & 0 & 	        p_1 & -p_2 & 0  &     p_1 & -p_2 & 0
\\ 
0 & p_1 & -p_1 & 			0 & 0 & 0 & 			0& p_3 & -p_2 
\\ 
-p_2 & 0 & p_2 & 			0 & 0 & 0 & 			 -p_3 & 0 & p_1
\\ 
 p_3 & -p_3 & 0 &			0 & 0 & 0 & 			 p_2 & -p_1 & 0
\\ 
0 & p_1 & -p_1 & 			0& p_3 & -p_2 	 & 0 & 0 & 0  			
\\ 
-p_2 & 0 & p_2 & 			-p_3 & 0 & p_1 	 & 0 & 0 & 0  		
\\ 
 p_3 & -p_3 & 0 &			p_2 & -p_1 & 0 	 & 0 & 0 & 0  		
\end{pmatrix} . 
\eeq
This can be written in block matrix form using the ${\bd X}$,  ${\bd Y}$,  ${\bd Z}$  matrices defined in eq. \rf{509},  
\beq{038}
P   =  R-R^t, 
\qquad \mbox{where} \quad
R = \begin{pmatrix} 
{\bd 0} & {\bd Y} &{\bd Y}  
\\ 
{\bd Z} &{\bd 0} & {\bd X}  
\\ 
{\bd Z}  & {\bd X}  &{\bd 0}
\end{pmatrix} . 
\eeq 
Note that this partition of the skew symmetric matrix is not unique.   

\subsection{Tensors of arbitrary order $n\ge2$}

The example of second order tensors shows that the three eigenvectors 
$\{ \ax ,\, {\bd v}_+,\, {\bd v}_- \}$ of the fundamental matrix ${\bd P}$ define all   $3^n=9 $ eigenvectors of $\tens P$.  This generalizes to arbitrary $n\ge 2$ by  generation of  all possible $n$-tensors formed from the tensor outer product of $\{ \ax ,\, {\bd v}_+,\, {\bd v}_- \}$.   
For any $n$, consider the product  $\tens P$  with the $n$-tensor 
\beq{041}
v_+ = \underbrace{{\bd v}_+ {\bd v}_+ \ldots {\bd v}_+}_{n}\, . 
\eeq
Each of the $n$ terms of $\tens P$ in \rf{594} contributes $+i v_+$ with the result
${\tens P} v_+ = ni\, v_+$.  Similarly, defining $v_-$ implies ${\tens P} v_- = -ni\, v_-$, from which it is clear that there are no other eigenvectors   with eigenvalues $\pm ni$.  Thus, the final component in the decomposition \rf{06a} is
\beq{043}
{\tens P}_n = i\, v_+ v_+^* - i\, v_- v_-^* \, . 
\eeq

Next consider the $n$ $n$-th order tensors that are combinations of  $(n-1)$ times ${\bd v}_+ $ with a single $\ax $, e.g. 
\beq{044}
v  = \ax\underbrace{{\bd v}_+ {\bd v}_+ \ldots {\bd v}_+}_{n-1}\, . 
\eeq
The action of ${\tens P}$  is  ${\tens P}v = (n-1)i v$, and so there are $2n$ eigenvectors with eigenvalues $\pm (n-1)i$ which  can be combined together to form the component ${\tens P}_{n-1}$, as in  eq. $\rf{0082}_1$ for  $n=2$.  Eigenvalues $ (n-2)i$ are obtained by combining $\ax$ twice with ${\bd v}_+$ $(n-2)$ times, but  also come from $n$-tensors formed by a single ${\bd v}_-$ with  $(n-1)$ times ${\bd v}_+$.  Enumeration yields a total of $n(n+1)$ eigenvectors associated with ${\tens P}_{n-2}$.  By recursion, it is clear that the number of eigenvectors with eigenvalues $ ki$  equals  the number of ways $n$ elements  chosen from $\{-1,0,1\}$ sum to $-n \le k\le n$, or equivalently, the coefficient of $x^k$ in the expansion of $\bigr(1+x+x^{-1}\bigr)^n$, 
\beq{964}
\bigr(1+x+x^{-1}\bigr)^n =  \sum\limits_{k=-n}^n\, \binom{n}{k}_2 \, x^k\, . 
\eeq
The numbers  ${\tbinom{n}{k}}_{_2}$ are   trinomial coefficients (not to be confused with $q$-multinomial coefficients \cite{Abramowitz74} with $q=3$)    and   can be expressed \cite{Andrews90} 
 \beq{9641}
\binom{n}{k}_2=  \sum\limits_{j=0}^n\, \binom{n}{2j+k}\binom{2j+k}{j}, 
\eeq
where ${\tbinom{n}{j}}  = n!/( j!(n-j)!)$ are binomial coefficients. 
 The main point for the present purpose is that
\beq{965}
 \sum\limits_{k=-n}^n\, \binom{n}{k}_2 = 3^n, 
\eeq
which is a consequence of \rf{964} with $x=1$. 
It is clear from  this  process that the entire set of $3^n$ eigenvectors has been obtained  and  there are no other candidates for eigenvalue of $\tens P$.   The precise form of the eigenvectors is unimportant because  the Euler-Rodrigues type formulas \rf{5191} depend only on the set of distinct non-zero eigenvalues.  However, the eigenvectors do play a role in setting up matrix representations of $\tens P$ for specific values of $n$, and in particular for tensors with symmetries, as discussed in Section \ref{apps}. 

\section{Applications to elasticity}\label{apps}

We first demonstrate that the underlying symmetry of  second and fourth order tensors in elasticity implies symmetries in the rotation tensors which reduces the number of independent elements in the latter.  

\subsection{Rotation of symmetric second and fourth order tensors}

We consider second and fourth order tensors such as the elastic  strain $\bd \varepsilon$ $(n=2)$ and  the  elastic stiffness tensor  $\tens C$ $(n=4)$ with elements $\varepsilon_{ij}$ and $C_{ijkl}$.  They transform as $\varepsilon_{ij} \rightarrow \varepsilon_{ij} '$ and $C_{ijkl} \rightarrow C_{ijkl}'$, 
\beq{qsat3}
\varepsilon_{ij} ' = Q_{ijpq}   \varepsilon_{pq} 
\qquad 
C_{ijkl}'= Q_{ijklpqrs}\, C_{pqrs}\, ,  
\eeq
where the  rotation tensors  follow from the general definition \rf{592} as
\beq{0503}
Q_{ijpq}  = Q_{ip}Q_{jq}, \qquad Q_{ijklpqrs} = Q_{ijpq} Q_{klrs} \, . 
\eeq
The strain and stiffness tensors possess physical symmetries which reduce the number of independent elements that need to be considered, 
\beq{053}
\varepsilon_{ij} = \varepsilon_{ji}, \qquad
C_{ijkl} = C_{jikl}= C_{ijlk}, \qquad
C_{ijkl} = C_{klij}. 
\eeq
In short, the number of independent elements of ${\bd \varepsilon}$ is reduced from $9$ to $6$, and of ${\tens C}$ from $81$ to $21$.  Accordingly, we may define variants of $Q_{ijkl}$ and $Q_{ijklpqrs}$ with a reduced number of elements that reflect the underlying symmetries of elasticity.  
Thus, the transformation rules can be expressed in the alternative forms 
\beq{qsat4}
\varepsilon_{ij} ' = \overline{Q}_{ijpq}  \varepsilon_{pq},
\qquad 
 C_{ijkl}' =   \overline{Q}_{ijklpqrs}\, C_{pqrs}\, , 
\eeq
where the elements of the symmetrized 4th and 8th order $\overline{\tens Q}$ tensors are 
\begin{subequations}
\bal{0050}
 \overline{Q}_{ijpq} &= \frac12   \big( Q_{ijpq} + Q_{ijqp}\big) ,
\\
 \overline{Q}_{ijklpqrs} &= \frac12   \big( 	\overline{Q}_{ijpq} \overline{Q}_{klrs} + 	\overline{Q}_{ijrs} \overline{Q}_{klpq}\big) \, . 
 \label{0050b}
\end{align}
Alternatively, they can be expressed in terms of the fundamental ${\bd Q} \in SO(3)$, 
\end{subequations}
\begin{subequations}
\bal{0051}
 \overline{Q}_{ijpq} &= \frac12   \big( Q_{ip}Q_{jq} + Q_{iq}Q_{jp}\big) ,
\\
 \overline{Q}_{ijklpqrs} 
  &=  \frac18   \big( 	
  Q_{ip}Q_{jq}Q_{kr}Q_{ls} +  Q_{ip}Q_{jq}Q_{ks}Q_{lr} + Q_{iq}Q_{jp}Q_{kr}Q_{ls} + 			Q_{iq}Q_{jp}Q_{ks}Q_{lr}
  \nonumber \\
 & \quad +
   Q_{ir}Q_{js}Q_{kp}Q_{lq} +  Q_{ir}Q_{js}Q_{kq}Q_{lp} + Q_{is}Q_{jr}Q_{kp}Q_{lq} + 			Q_{is}Q_{jr}Q_{kq}Q_{lp}
  \big)
 \, .
\label{0051b}
\end{align}
\end{subequations}
Using the identity \rf{600}, we have 
\beq{630}
\overline{Q}_{ijpq}\overline{Q}_{klpq} =  \frac12 \big( I_{ijkl} +I_{ijlk} \big) = 
\overline{I}_{ijkl}
\equiv \frac12 \big( \delta_{ik}\delta_{jl}+ \delta_{il}\delta_{jk} \big), 
\eeq
which is the fourth order isotropic identity tensor, i.e. the fourth order tensor with the property $s=\overline{\tens I}s$ for all symmetric second order tensors $s$. Similarly, 
\beq{631}
\overline{Q}_{ijklabcd}\overline{Q}_{pqrsabcd} =  \frac12 \big( 
\overline{I}_{ijpq}\overline{I}_{klrs} +\overline{I}_{ijrs}\overline{I}_{klpq}
 \big) \equiv 
\overline{I}_{ijklpqrs}\, .  
\eeq
This is the 8th order isotropic tensor with the property ${\tens C}=\overline{\tens I}{\tens C}$ for all 4th order elasticity tensors.  In summary, the symmetrized rotation tensors for strain and elastic moduli are orthogonal in the sense that  $\overline{\tens Q}\, \overline{\tens Q}^t = \overline{\tens I}$, and are therefore elements of $SO(6)$ and $SO(21)$, respectively. 

The (symmetrized) rotation tensor for strain  displays the following symmetries
\beq{0052}
\overline{Q}_{ijpq} = \overline{Q}_{jipq} = \overline{Q}_{ijqp}.  
\eeq
Introducing the  Voigt indices, which are  capital suffices taking the values $1,2,\ldots, 6$ according to 
\beq{904}
I=1,2,3,4,5, 6 \qquad \Leftrightarrow \qquad  ij = 11,22,33,23,31,12 , 
\eeq
then \rf{0052} implies that the elements $\overline{Q}_{ijpq}$ can be represented   by  $\overline{Q}_{IJ}$.  Similarly,  the elements of the 8th order tensor $\overline{Q}_{ijklpqrs}$ can be represented as $\overline{Q}_{IJKL}$, which  satisfy the symmetries
\beq{0053}
\overline{Q}_{IJKL} = \overline{Q}_{JIKL} = \overline{Q}_{IJLK}.  
\eeq
The pairs of indices $IJ$ and $KL$ each represent $21$ independent values, suggesting the introduction of a new type of suffix which ranges from $1,2,\ldots , 21$. Thus, 
\bal{905}
{\cal I}&=1,2,3,4,5, 6,7,8,9,10,11,12,13,14,15,16,17,18,19,20,21
\nonumber \\ 
\qquad \Leftrightarrow \qquad  IJ &= 11,22,33,23,31,12, 44,55,66,14,25,36,34,15,26,24,35,16,56,64,45. 
\end{align}
Hence  the elements $\overline{Q}_{ijklpqrs}$ can be represented uniquely by the  $(21)^2$ elements 
$\overline{Q}_{\cal IJ}$.  The elements of the  symmetrized rotations $\overline{Q}_{IJ}$ and $\overline{Q}_{\cal IJ}$ have reduced dimensions, $6$ and $21$ respectively, but do not yet represent the matrix elements of $6$- and $21$-dimensional tensors.  That step is completed next, after which we can define the associated  skew symmetric generating  matrices and return to the 
 generalized Euler-Rodrigues formulas for  $6$- and $21$-dimensional matrices.

\subsection{Six dimensional representation}

We now use the isomorphism between second order tensors in three dimensions, such as the strain $\boldsymbol{\varepsilon}$, and six dimensional vectors according to $\boldsymbol{\varepsilon} \rightarrow \hat{\boldsymbol\varepsilon}$ with elements $\hat{\varepsilon}_I$, $I=1,2,\ldots, 6$.  Similarly, 
fourth order elasticity tensors in three dimensions are  isomorphic with second order positive definite symmetric tensors in six dimensions $\whbf{C}$ with elements $  \hat{c}_{IJ}$ \cite{c3}.   
Let $\{ \varepsilon \}$ be the $6$-vector with elements $\varepsilon_I$, $I=1,2,\ldots , 6$, 
and $[C]$ the $6\times 6$ Voigt matrix of elastic moduli, i.e. with elements $c_{IJ}$.  The associated $6$-dimensional vector and tensor are 
\beq{defs}
\hat{\boldsymbol\varepsilon}  = {\bf T} \{ \varepsilon \},
\qquad
\whbf{C} = {\bf T}   [C]  {\bf T},
\quad \mbox{where  }
{\bf T} \equiv 
{\diag} \big(1,\,1,\,1,\,\sqrt{2},\,\sqrt{2},\,\sqrt{2} \big)   .
\eeq
Explicitly,   
\beq{a1}
\hat{\boldsymbol\varepsilon} 
=  \begin{pmatrix}
\hat\varepsilon_{1}
\\  
\hat\varepsilon_{2}
\\  
\hat\varepsilon_{3}
\\  
\hat\varepsilon_{4}
\\  
\hat\varepsilon_{5}
\\  
\hat\varepsilon_{6}
\end{pmatrix}=  \begin{pmatrix}
\varepsilon_{11}
\\  
\varepsilon_{22}
\\  
\varepsilon_{33}
\\  
\sqrt{2} \varepsilon_{23}
\\  
\sqrt{2} \varepsilon_{31}
\\  
\sqrt{2} \varepsilon_{12}
\end{pmatrix}, \qquad \quad \whbf{C}   =  \begin{pmatrix}
c_{11} & c_{12} & c_{13} & 
 2^{\frac12} c_{14} & 2^{\frac12} c_{15} & 2^{\frac12} c_{16} 
\\ 
 & c_{22} & c_{23} & 
 2^{\frac12} c_{24} & 2^{\frac12} c_{25} & 2^{\frac12} c_{26} 
\\ 
 &  & c_{33} & 
 2^{\frac12} c_{34} & 2^{\frac12} c_{35} & 2^{\frac12} c_{36} 
\\ 
 &  &  & 2c_{44} & 2c_{45} & 2c_{46}
\\ 
 {\rm S} & {\rm Y} & {\rm M} & & 2c_{55} & 2c_{56}
\\ 
& & & & & 2c_{66}
\end{pmatrix} \, .  
\eeq
The $\sqrt{2}$ terms ensure that products and norms are preserved, e.g. 
$C_{ijkl}C_{ijkl} = \tr \whbf{C}^t\whbf{C}$. 

The six-dimensional version of the fourth order tensor $\overline{Q}_{ijkl}$ is  
$\whbf{Q} \in SO(6)$,  introduced by Mehrabadi et al. \cite{mcj}, see also \cite{Norris05}.  It may be defined in the same manner as $\whbf{Q} = {\bd S}[\overline{Q}]{\bd S}$ where $[\overline{Q}]$ is the matrix of Voigt elements.   
Thus, $ \whbf{Q}\whbf{Q}^t = \whbf{Q}^t\whbf{Q} = \whbf{I}$, 
 where $\whbf{I} = \diag (1,1,1,1,1,1)$.   It can be expressed  
$ \whbf{Q} (\ax , \theta ) =  \exp (\theta  \whbf{P}) $,
and hence is given by the $n=2$ Euler-Rodrigues formula \rf{14} for  the skew symmetric  $\whbf{P}(\ax) \in so(6)$, 
\beq{0250}
\whbf{P}  =  \begin{pmatrix} 
0 & 0 & 0 & 
        0 & \sqrt{2} p_2 & -\sqrt{2} p_3  
\\ 
0 & 0 & 0 & 
        -\sqrt{2} p_1 &0 &  \sqrt{2} p_3
\\ 
0 & 0 & 0 & 
        \sqrt{2} p_1 &  -\sqrt{2} p_2 &0
\\ 
0 & \sqrt{2} p_1 & -\sqrt{2} p_1   & 0 & p_3 & -p_2
\\ 
-\sqrt{2} p_2 &0 &  \sqrt{2} p_2  & -p_3 & 0 & p_1
\\ 
\sqrt{2} p_3 &  -\sqrt{2} p_3 &0 &p_2 & -p_1 & 0 
\end{pmatrix} , 
\eeq
or in terms of  the block matrices  defined in \rf{509},
\beq{0251}
\whbf{P} = \whbf{R}- \whbf{R}^t,\qquad
\whbf{R}  =  \begin{pmatrix}
{\bf 0} & \sqrt{2}{\bf Y} 
       \\ 
\sqrt{2}{\bf Z} & {\bf X}  
\end{pmatrix} .
\eeq
It is useful to compare $\whbf{P}$ with the  $9\times 9$ matrix for rotation of general second order tensors, eqs. \rf{037} and \rf{038}.  The reduced dimensions  and the $\sqrt{2}$ terms  are a consequence of the underlying symmetry of the tensors that are being rotated.  Vectors and tensors transform as 
$\hat{\boldsymbol\varepsilon}  \rightarrow  \hat{\boldsymbol\varepsilon}'$ 
and $\whbf{C} \rightarrow \whbf{C} '$ where
\beq{t2}
\hat{\boldsymbol\varepsilon}' = \whbf{Q} \, \hat{\boldsymbol\varepsilon}, \qquad 
 \whbf{C} ' = \whbf{Q}\whbf{C}\whbf{Q}^t \, . 
\eeq

The eigenvalues and orthonormal eigenvectors of $\whbf{P}$  are as follows
\beq{ev6}
\begin{matrix}
\mbox{eigenvalue}  & \qquad \mbox{eigenvector} \qquad  & \qquad   \mbox{dyadic}  \qquad 
\\ 
 0 & \hat{\bf i} & \frac1{\sqrt{3}}{\bf I} 
\\ 
  0 & \hat\ax_{\rm d}  & \sqrt{\frac32}(  \ax  \ax - \frac13{\bf I}) 
\\ 
 \pm i &  \hat{\bf u}_\pm  & \frac1{\sqrt{2}}(\ax  {\bf v}_\pm +{\bf v}_\pm   \ax) 
\\ 
  \pm 2i & \hat{\bf v}_\pm & {\bf v}_\pm   {\bf v}_\pm
\end{matrix}
\eeq
where ${\bf v}_\pm$ are defined in \rf{ev3} and 
\beq{925}
\hat{\bf i} \equiv 
\begin{pmatrix} 1/\sqrt{3} \\  1/\sqrt{3} \\  1/\sqrt{3} \\ 0 \\   0\\ 0 \end{pmatrix}
\quad
 \hat\ax_{\rm d} \equiv \sqrt{\frac32}
\begin{pmatrix} p_1^2 -\frac13 \\  p_2^2 -\frac13\\  p_3^2 -\frac13\\ \sqrt{2}\,  p_2p_3 \\   
			\sqrt{2}\,  p_3p_1\\ \sqrt{2}\,  p_1p_2 \end{pmatrix},
\quad 
  \hat{\bf u}_\pm \equiv 
  \begin{pmatrix} \sqrt{2}\, p_1 v_{\pm,1} \\ \sqrt{2}\, p_2 v_{\pm,2} \\ \sqrt{2}\,  p_3 v_{\pm,3}   \\
  p_2 v_{\pm,3}+p_3 v_{\pm,2} \\
p_3 v_{\pm,1}+p_1 v_{\pm,3} \\
 p_1 v_{\pm,2}+p_2 v_{\pm,1} 
   \end{pmatrix},
\quad 
  \hat{\bf v}_\pm \equiv 
\begin{pmatrix} v_{\pm,1}^2  \\ v_{\pm,2}^2 \\ v_{\pm,3}^2 \\ 
	\sqrt{2}\,  v_{\pm,2} v_{\pm,3}\\ \sqrt{2}\,  v_{\pm,3} v_{\pm,1}\\
		\sqrt{2}\,  v_{\pm,1} v_{\pm,2}\end{pmatrix}\, . 
\eeq
The right column in \rf{ev6} shows the  second order tensors corresponding to the eigenvectors in dyadic form. 
The eigenvectors are orthonormal, i.e. of unit magnitude and mutually orthogonal. The unit $6-$vectors $\hat{\bf i}$ and $\hat\ax_{\rm d}$ correspond to the hydrostatic and deviatoric parts of  $\ax\ax $, respectively. 
The double multiplicity of the zero eigenvalue means that any $6-$vector of the form $ a\hat\ax_{\rm d}+b\hat{\bf i}$ is a null vector of $\whbf{P}$, including 
$ \sqrt{\tfrac23}
\hat\ax_{\rm d} + \tfrac1{\sqrt{3}}\hat{\bf i}$ which is the six-vector for the dyad $\ax \ax$. 

The canonical decomposition of the generating matrix $\whbf{P}\in so(6)$ is therefore
\beq{p13}
\whbf{P} = \whbf{P}_1+2\whbf{P}_2,\qquad
\whbf{P}_1= 
i  \hat{\bf u}_+ \hat{\bf u}_+^* - i  \hat{\bf u}_-\hat{\bf u}_-^*, 
\quad
\whbf{P}_2= 
i  \hat{\bf v}_+ \hat{\bf v}_+^* - i  \hat{\bf v}_-\hat{\bf v}_-^*\, . 
 \eeq
The associated   fourth order tensor $\overline{\tens P}$, which is the symmetrized version of $\tens P$ in eq. \rf{0081}, is 
\beq{p16}
\overline{\tens P} = \overline{\tens P}_1+2\overline{\tens P}_2,
\qquad
\overline{\tens P}_1 = 
\sum\limits_{\pm} \, 
\frac{i}2 (\ax  {\bf v}_\pm +{\bf v}_\pm   \ax)
  (\ax  {\bf v}_\pm^* +{\bf v}_\pm^*   \ax),
  \quad
  \overline{\tens P}_2 = 
\sum\limits_{\pm} \, \pm i\,  {\bf v}_\pm   {\bf v}_\pm   {\bf v}_\pm^*  {\bf v}_\pm^* .
 \eeq

The representation \rf{p13} allows us to compute powers of $ \whbf{P}$, and using the spectral decomposition of the identity 
\beq{p133}
\whbf{I} = \hat{\bf i}\hat{\bf i}^t + \hat\ax_{\rm d}\hat\ax_{\rm d}^t + \hat{\bf u}_+ \hat{\bf u}_+^*  +  \hat{\bf u}_-\hat{\bf u}_-^*
+ \hat{\bf v}_+ \hat{\bf v}_+^* +  \hat{\bf v}_-\hat{\bf v}_-^*\, , 
 \eeq
the characteristic equation \cite{mcj} follows: 
\beq{206}
\whbf{P} (\whbf{P}^2 +\whbf{I})
(\whbf{P}^2 +4\whbf{I}) = 0 \, . 
 \eeq

\subsection{21 dimensional representation}

Vectors and matrices in 21-dimensions are denoted by a tilde.  Thus, the 21-dimensional vector of elastic moduli is $\tilde{\bf c} $ with elements 
$\tilde{c}_{\cal I}$, ${\cal I}=1,2,\ldots, 21$.  Similarly, 
 vectors corresponding to the fourth order tensors $\boldsymbol{\varepsilon} \boldsymbol{\varepsilon}$ and $\frac12({\bf A}\otimes {\bf B}+{\bf B}\otimes {\bf A})$  are $\tilde{\boldsymbol{\varepsilon}}$ and $\widetilde{\bf v}$, that is 
 \beq{211}
 \begin{matrix}
\mbox{4$^{th}$ order, n=3}  & \qquad \mbox{2$^{nd}$ order, n=6} \qquad &   \mbox{vector, n=21}  
\\ 
\tens{C} & \whbf{C} & \tilde{\bf c}
\\ 
\boldsymbol{\varepsilon} \boldsymbol{\varepsilon} &  \hat{\boldsymbol{\varepsilon}}\hat{\boldsymbol{\varepsilon}}^t 		
		& \tilde{\boldsymbol{\varepsilon}} 
		\\ 
~~ \frac12({\bf A}\otimes {\bf B}+{\bf B}\otimes {\bf A}) ~~ & 
~~ \frac12(\hat{\bf a} \hat{\bf b}^t +\hat{\bf b} \hat{\bf a}^t  ) ~~
 			& \widetilde{\bf v}
\end{matrix}  \, . 
 \eeq
The $21$ elements follow from the indexing scheme \rf{905} and are  defined by 
 { \fontsize{11}{15}   \selectfont
\begin{align}\label{t32}
\tilde{\bf c} = \begin{pmatrix} \tilde{c}_1 \\ \tilde{c}_2 \\ 
\tilde{c}_3 \\ 
\tilde{c}_4 \\ 
\tilde{c}_5 \\ 
\tilde{c}_6 \\ \tilde{c}_7 \\ 
\tilde{c}_8 \\ 
\tilde{c}_9 \\ 
\tilde{c}_{10} \\ 
\tilde{c}_{11} \\ 
\tilde{c}_{12} \\ 
\tilde{c}_{13} \\ 
\tilde{c}_{14} \\ 
\tilde{c}_{15} \\ 
\tilde{c}_{16} \\ 
\tilde{c}_{17} \\ 
\tilde{c}_{18} \\ 
\tilde{c}_{19} \\ 
\tilde{c}_{20} \\ 
\tilde{c}_{21}
\end{pmatrix}
=\begin{pmatrix}
c_{11} \\ 
c_{22} \\ 
c_{33} \\   
\sqrt{2}c_{23} \\ 
\sqrt{2}c_{13} \\ 
\sqrt{2}c_{12} \\ 
 2c_{44} \\ 
 2c_{55} \\ 
 2c_{66} \\ 
 2c_{14} \\ 
 2c_{25} \\ 
 2c_{36} \\ 
  2c_{34} \\ 
  2c_{15} \\ 
  2c_{26} \\ 
  2c_{24} \\ 
  2c_{35} \\ 
  2c_{16} \\ 
  2\sqrt{2}c_{56} \\ 
  2\sqrt{2}c_{46} \\ 
  2\sqrt{2}c_{45}
  \end{pmatrix}
 =\begin{pmatrix}
\hat{c}_{11} \\ 
\hat{c}_{22} \\ 
\hat{c}_{33} \\ 
\sqrt{2}\hat{c}_{23} \\ 
\sqrt{2}\hat{c}_{13} \\ 
\sqrt{2}\hat{c}_{12} \\ 
\hat{c}_{44} \\ 
\hat{c}_{55} \\ 
\hat{c}_{66} \\ 
\sqrt{2}\hat{c}_{14} \\ 
\sqrt{2}\hat{c}_{25} \\ 
\sqrt{2}\hat{c}_{36} \\ 
\sqrt{2}\hat{c}_{34} \\ 
\sqrt{2}\hat{c}_{15} \\ 
\sqrt{2}\hat{c}_{26} \\ 
\sqrt{2}\hat{c}_{24} \\ 
\sqrt{2}\hat{c}_{35} \\ 
\sqrt{2}\hat{c}_{16} \\ 
\sqrt{2}\hat{c}_{56} \\ 
\sqrt{2}\hat{c}_{46} \\ 
\sqrt{2}\hat{c}_{45}
\end{pmatrix}, 
\qquad
  \tilde{\boldsymbol{\varepsilon}} = 
  \begin{pmatrix}
  \varepsilon_{11}^2 \\
  \varepsilon_{22}^2 \\ 
  \varepsilon_{33}^2 \\ 
\sqrt{2}\varepsilon_{22}\varepsilon_{33} \\ 
\sqrt{2}\varepsilon_{11}\varepsilon_{33} \\ 
\sqrt{2}\varepsilon_{11}\varepsilon_{22} \\ 
  2\varepsilon_{23}^2 \\ 
  2\varepsilon_{13}^2 \\ 
  2\varepsilon_{12}^2 \\ 
  2\varepsilon_{11}\varepsilon_{23} \\  
  2\varepsilon_{22}\varepsilon_{13} \\ 
  2\varepsilon_{33}\varepsilon_{12} \\   
  2\varepsilon_{33}\varepsilon_{23} \\  
  2\varepsilon_{11}\varepsilon_{13} \\   
  2\varepsilon_{22}\varepsilon_{12} \\   
  2\varepsilon_{22}\varepsilon_{23} \\ 
  2\varepsilon_{33}\varepsilon_{13} \\ 
  2\varepsilon_{11}\varepsilon_{12} \\ 
  2\sqrt{2}\varepsilon_{13}\varepsilon_{12} \\    	 		      2\sqrt{2}\varepsilon_{23}\varepsilon_{12} \\ 
   2\sqrt{2}\varepsilon_{23}\varepsilon_{13}
   \end{pmatrix}
   =  \begin{pmatrix}
   \hat\varepsilon_{1}^2 \\ 
   \hat\varepsilon_{2}^2 \\ \hat\varepsilon_{3}^2 \\ 
\sqrt{2}\hat\varepsilon_{2}\hat\varepsilon_{3} \\ 
\sqrt{2}\hat\varepsilon_{3}\hat\varepsilon_{1} \\ 
\sqrt{2}\hat\varepsilon_{1}\hat\varepsilon_{2} \\ 
  \hat\varepsilon_{4}^2 \\ 
  \hat\varepsilon_{5}^2 \\ 
  \hat\varepsilon_{6}^2 \\ 
  \sqrt{2}\hat\varepsilon_{1}\hat\varepsilon_{4} \\  \sqrt{2}\hat\varepsilon_{2}\hat\varepsilon_{5} \\ 
 \sqrt{2}\hat\varepsilon_{3}\hat\varepsilon_{6} \\   
   \sqrt{2}\hat\varepsilon_{3}\hat\varepsilon_{4} \\  \sqrt{2}\hat\varepsilon_{1}\hat\varepsilon_{5} \\   
  \sqrt{2}\hat\varepsilon_{2}\hat\varepsilon_{6} \\    \sqrt{2}\hat\varepsilon_{2}\varepsilon_{4} \\ 
  \sqrt{2}\hat\varepsilon_{3}\hat\varepsilon_{5} \\ 
  \sqrt{2}\hat\varepsilon_{1}\hat\varepsilon_{6} \\ 
  \sqrt{2}\hat\varepsilon_{5}\hat\varepsilon_{6} \\  \sqrt{2}\hat\varepsilon_{6}\hat\varepsilon_{4} \\ 
   \sqrt{2}\hat\varepsilon_{4}\hat\varepsilon_{5}
      \end{pmatrix}, \qquad 
  \widetilde{\bf v} = 
  \begin{pmatrix}
  \hat{a}_1\hat{b}_1 \\
  \hat{a}_2\hat{b}_2 \\
  \hat{a}_3\hat{b}_3 \\
  \frac1{\sqrt{2}}(\hat{a}_2\hat{b}_3+\hat{a}_3\hat{b}_2)\\
	\frac1{\sqrt{2}}(\hat{a}_3\hat{b}_1+\hat{a}_1\hat{b}_3)\\
	\frac1{\sqrt{2}}(\hat{a}_1\hat{b}_2+\hat{a}_2\hat{b}_1)\\	
	\hat{a}_4\hat{b}_4 \\
	\hat{a}_5\hat{b}_5 \\
	\hat{a}_6\hat{b}_6 \\
	\frac1{\sqrt{2}}(\hat{a}_1\hat{b}_4+\hat{a}_4\hat{b}_1)\\
	\frac1{\sqrt{2}}(\hat{a}_2\hat{b}_5+\hat{a}_5\hat{b}_2)\\
	\frac1{\sqrt{2}}(\hat{a}_3\hat{b}_6+\hat{a}_6\hat{b}_3)\\	
  \frac1{\sqrt{2}}(\hat{a}_3\hat{b}_4+\hat{a}_4\hat{b}_3)\\
	\frac1{\sqrt{2}}(\hat{a}_1\hat{b}_5+\hat{a}_5\hat{b}_1)\\
	\frac1{\sqrt{2}}(\hat{a}_2\hat{b}_6+\hat{a}_6\hat{b}_2)\\	
	\frac1{\sqrt{2}}(\hat{a}_2\hat{b}_4+\hat{a}_4\hat{b}_2)\\
	\frac1{\sqrt{2}}(\hat{a}_3\hat{b}_5+\hat{a}_5\hat{b}_3)\\
	\frac1{\sqrt{2}}(\hat{a}_1\hat{b}_6+\hat{a}_6\hat{b}_1)\\	
	\frac1{\sqrt{2}}(\hat{a}_5\hat{b}_6+\hat{a}_6\hat{b}_5)\\
	\frac1{\sqrt{2}}(\hat{a}_6\hat{b}_4+\hat{a}_4\hat{b}_6)\\
	\frac1{\sqrt{2}}(\hat{a}_4\hat{b}_5+\hat{a}_5\hat{b}_4)
 \end{pmatrix}	
\, . \qquad \qquad 
\end{align}
}
Alternatively, 
\beq{572}
\tilde{\bf c}  = \widetilde{\bd T} \{ c\}, \quad {\rm etc.}, 
\eeq
where $\{ c\}$ is the $21\times 1$ array with elements $c_{\cal I}$, and $\widetilde{\bd T}$ is 
the the $21\times 21$ diagonal with block structure
\beq{573}   
\widetilde{\bd T}  = \diag \big( {\bd I}\ \sqrt{2}{\bd I}\ \ 
2 {\bd I}\ \ 2 {\bd I}\ \ 2 {\bd I}\ \ 2 {\bd I}\ \ 2 \sqrt{2} {\bd I}\big)\, . 
\eeq
For instance, the  elastic energy density $W= \tfrac12 C_{ijkl}\varepsilon_{ij}\varepsilon_{kl}$ is  given by the vector inner product 
\beq{804}
W = \frac12\, \widetilde{\boldsymbol{\varepsilon}}^t  \widetilde{\bf c}\, , 
\eeq
where $\widetilde{\boldsymbol{\varepsilon}}$ is the $21$-vector associated with strain  \rf{t32}.

\subsubsection{The 21-dimensional rotation matrix} 

Vectors in 21-dimensions  transform as 
$ \tilde{\bf c}  \rightarrow  \tilde{\bf c}'= \widetilde{\bf Q}\, \tilde{\bf c}$ 
where  $ \widetilde{\bf Q} \in $SO(21) satisfies 
$
\widetilde{\bf Q}\widetilde{\bf Q}^t = \widetilde{\bf Q}^t\widetilde{\bf Q} = \widetilde{\bf I}$, 
 and $\widetilde{\bf I} $ is the identity, $\widetilde{I}_{\cal IJ}= \delta_{\cal IJ}$. 
The rotation matrix $\widetilde{\bf Q}$ can be expressed 
\bal{68} 
\widetilde{\bf Q}  (\ax , \theta ) = e^{ \theta \widetilde{\bf P} }, 
\end{align}
where $\widetilde{\bf P}(\ax)\in so(21)$ follows from the general definition \rf{594} applied to elasticity tensors, i.e. \rf{0051b}.  Thus, 
\beq{575}
\widetilde{\bf P}  = \widetilde{\bd T} [ \overline{P} ] \widetilde{\bd T}, 
\qquad
 \widetilde{\bf Q}  = \widetilde{\bd T} [ \overline{Q} ] \widetilde{\bd T}
\eeq
where $[ \overline{P} ], \, [ \overline{Q} ]$ are  the $21\times 21$ array with elements $\overline{P}_{\cal IJ},\, \overline{Q}_{\cal IJ}$.  We take a slightly different approach, and derive the elements of $\widetilde{\bf P}$ using the 
expansion of $\widetilde{\bf Q}$ for small $\theta$ directly.  Define the rotational derivative of the moduli  as
\beq{69}
\dot{c}_{ijkl} =  \left. \frac{d c'_{ijkl} }{d\theta}\right|_{\theta = 0}. 
\eeq
Using   $\rf{t2}_2$ for instance, gives  
\beq{s8}
 \dot{\whbf{C}}  =  \whbf{P}  \whbf{C}   + \whbf{C}  \whbf{P}^t \, ,  
\eeq
from which we obtain 
\begin{subequations} \label{map}
\bal{mapa}
  \dot{c}_{11} &= 4 c_{15} p_2 - 4 c_{16} p_3 
 \, , \\  \dot{c}_{22} &=   -4  c_{24} p_1 + 4 c_{26} p_3 
 \, , \\  \dot{c}_{33} &=     4 c_{34} p_1 - 4 c_{35} p_2 
 \, , \\  \dot{c}_{23} &=   2 (c_{24} -  c_{34}) p_1 - 2 c_{25} p_2  + 2 c_{36} p_3  
 \, , \\  \dot{c}_{13} &=    2 c_{14} p_1 + 2 (c_{35} -  c_{15}) p_2 - 2 c_{36} p_3  
 \, , \\  \dot{c}_{12} &=   -2 c_{14} p_1  + 2 c_{25} p_2 + 2 (c_{16} -  c_{26} ) p_3 
 \, , \\  \dot{c}_{44} &=    2(c_{24} -  c_{34} )p_1   - 2 c_{46} p_2   + 2 c_{45} p_3  
 \, , \\  \dot{c}_{55} &=  2 c_{56} p_1  +2 (c_{35} - c_{15}) p_2 - 2 c_{45} p_3  
 \, , \\  \dot{c}_{66} &=    -2 c_{56} p_1 + 2 c_{46} p_2   + 2 (c_{16} -  c_{26} )p_3
 \, , \\  \dot{c}_{14} &=  ( c_{12}  - c_{13}) p_1  - ( c_{16} -2 c_{45}   ) p_2 + ( c_{15}- 2 c_{46} )p_3
 \, , \\  \dot{c}_{25} &=  (c_{26}-2 c_{45} )p_1 +  (c_{23}  - c_{12} )p_2 - (c_{24}   - 2 c_{56} )p_3  
 \, , \\  \dot{c}_{36} &=   -(c_{35}  - 2 c_{46} )p_1  + (c_{34} -2 c_{56} )p_2  + (c_{13} - c_{23}) p_3 
 \, , \\  \dot{c}_{34} &=    -( c_{33} - c_{23} - 2 c_{44} )p_1 - (c_{36}   + 2 c_{45} )p_2  +  c_{35} p_3
 \, , \\  \dot{c}_{15} &=   c_{16} p_1 - ( c_{11}   - c_{13} - 2 c_{55}) p_2 - (c_{14}    + 2 c_{56}) p_3  
 \, , \\  \dot{c}_{26} &=   -(c_{25} + 2 c_{46}) p_1 + c_{24} p_2 - ( c_{22}- c_{12}    - 2 c_{66} )p_3 
 \, , \\  \dot{c}_{24} &=  ( c_{22}- c_{23} - 2 c_{44} ) p_1 - c_{26} p_2  + ( c_{25}+2 c_{46} )   p_3  
 \, , \\  \dot{c}_{35} &=   (c_{36}  + 2 c_{45}) p_1 +( c_{33}  - c_{13}- 2 c_{55} ) p_2 - c_{34} p_3      
 \, , \\  \dot{c}_{16} &=    -c_{15} p_1 + (c_{14} + 2 c_{56}) p_2  + (c_{11} - c_{12}    - 2 c_{66} )p_3 
 \, , \\  \dot{c}_{56} &=    (c_{66} - c_{55} )p_1 + (c_{36} -c_{16} + c_{45} ) p_2  - (c_{25} - c_{15} + c_{46}) p_3
 \, , \\  \dot{c}_{46} &=   -(c_{36}  - c_{26}   + c_{45} ) p_1 + ( c_{44} - c_{66} ) p_2   + (c_{14} - c_{24} + c_{56}) p_3 
 \, , \\  \dot{c}_{45} &=  ( c_{25} - c_{35} + c_{46} ) p_1 - (  c_{14}  - c_{34} + c_{56}) p_2 + (c_{55} - c_{44} )p_3  
  \, . \quad
\end{align}
\end{subequations}
The elements of $\widetilde{\bf P}$ can be read off from this and are given in Table 1.   
The $21\times 21$ format  can be simplified to a $7\times 7$ array of block elements, comprising combinations of the $3\times 3$ zero matrix $\bf 0$ and the  $3\times 3$ matrices defined in \rf{509}.  Thus, 
\begin{align}\label{213}
\widetilde{\bf P} = \widetilde{\bf R} -\widetilde{\bf R}^t, \quad \mbox{where} \quad 
\widetilde{\bf R} & =  \begin{pmatrix} 
{\bf 0} & {\bf 0} & {\bf 0} & {\bf 0} &2{\bf Y}  & {\bf 0} &{\bf 0} 
					\\ 
{\bf 0} & {\bf 0} & {\bf 0} & -\sqrt{2}{\bf Y}  & {\bf 0}  & \sqrt{2}{\bf N}  &{\bf 0} 
					\\ 
{\bf 0} & {\bf 0} & {\bf 0} & {\bf 0} & {\bf 0} & 2{\bf N}  & -\sqrt{2}{\bf Y}  
					\\ 
{\bf 0} & -\sqrt{2}{\bf Z} & {\bf 0} & {\bf 0} & {\bf X} &{\bf 0} &  -\sqrt{2}{\bf X}  
					\\ 
{\bf 0}  &\sqrt{2}{\bf N} & 2{\bf N} & {\bf 0} & {\bf 0} & {\bf X}  & {\bf 0}
					\\ 
 2{\bf Z}   & {\bf 0} & {\bf 0} &   {\bf X} & {\bf 0} & {\bf 0} &   \sqrt{2}{\bf X}
   				\\ 
{\bf 0} &{\bf 0} & -\sqrt{2}{\bf Z}  & - \sqrt{2}{\bf X}   &\sqrt{2}{\bf X} & {\bf 0} &  - {\bf X}
\end{pmatrix}
\, . 
\end{align}
Equation \rf{2133} shows that the rotation matrices in three, six and 21 dimensions can  be simplified using the fundamental  $3\times 3$ matrices $ {\bf X}, {\bf Y}, {\bf Z}$ and $ {\bf N}$.  The partition of a skew symmetric matrix in this way is not unique, but it simplifies the calculation and numerical implementation.  These  matrices satisfy several  identities, 
\begin{subequations}
\begin{align}
&\qquad{\bf X}{\bf X}^t + {\bf Y}{\bf Y}^t + {\bf Z}{\bf Z}^t = {\bf I}\, , 
\\
& {\bf X}^t{\bf X} = {\bf Y}{\bf Y}^t,\quad 
{\bf Y}^t{\bf Y} = {\bf Z}{\bf Z}^t,\quad 
{\bf Z}^t{\bf Z} = {\bf X}{\bf X}^t,\quad 
\\
& {\bf X}^t{\bf Y} = {\bf Y}{\bf Z}^t,\quad 
{\bf Y}^t{\bf Z} = {\bf Z}{\bf X}^t,\quad 
{\bf Z}^t{\bf X} = {\bf X}{\bf Y}^t,\quad 
\\
&  \|{\bf N}\| = \|{\bf X}\| = \|{\bf Y}\| = \|{\bf Z}\| =   1,  
\end{align}
\end{subequations}
 where the norm is    $\| u \|  = \sqrt{ {\tr}\, u^t \, u }$ .

We find that the eigenvalues of $\widetilde{\bf P}$ are $0 (5),\, i (3),\, -i (3),\, 2i (3),\, -2i (3),\, 3i,\, 
- 3i,\,  4i,\, -4i$, where the number in parenthesis is the multiplicity.  
The associated  orthonormal eigenvectors of $\widetilde{\bf P}$ are 
\beq{ev21}
\begin{matrix}
\mbox{eigenvalue}  & \qquad \mbox{eigenvector}  \qquad  &   \mbox{6-dyadic} 
\\  \\ 
0& \widetilde{\bf v}_{0a} & \hat{\bf i}\hat{\bf i}^t  
\\ 
0& \widetilde{\bf v}_{0b} & \frac32\hat\ax_{\rm d}\hat\ax_{\rm d}^t  
\\ 
0& \widetilde{\bf v}_{0c} & \frac{\sqrt{3}}{2}(\hat{\bf i}\hat\ax_{\rm d}^t + \hat\ax_{\rm d}\hat{\bf i}^t)   
\\ 
0& \widetilde{\bf v}_{0d} & \frac1{\sqrt{2}}(\hat{\bf u}_- \hat{\bf u}_+^t + \hat{\bf u}_+ \hat{\bf u}_-^t)  
\\ 
0& \widetilde{\bf v}_{0e} & \frac1{\sqrt{2}}(\hat{\bf v}_- \hat{\bf v}_+^t + \hat{\bf v}_+ \hat{\bf v}_-^t)  
\\ 
\pm i  & \widetilde{\bf v}_{1a\pm} & \frac1{\sqrt{2}}(\hat{\bf i}\hat{\bf u}_\pm^t + \hat{\bf u}_\pm\hat{\bf i}^t)   
\\ 
\pm i  & \widetilde{\bf v}_{1b\pm} & \frac{\sqrt{3}}{2}(\hat\ax_{\rm d}\hat{\bf u}_\pm^t + \hat{\bf u}_\pm\hat\ax_{\rm d}^t)   
\\ 
\pm i  & \widetilde{\bf v}_{1c\pm} & \frac1{\sqrt{2}}(\hat{\bf v}_\pm \hat{\bf u}_\mp^t + \hat{\bf u}_\mp \hat{\bf v}_\pm^t)   
\\ 
\pm 2i &    \widetilde{\bf v}_{2a\pm } &   \hat{\bf u}_\pm \hat{\bf u}_\pm^t    
\\ 
\pm 2i &    \widetilde{\bf v}_{2b\pm } &  \frac1{\sqrt{2}}(\hat{\bf i}\hat{\bf v}_\pm^t + \hat{\bf v}_\pm\hat{\bf i}^t)  
\\ 
\pm 2i &    \widetilde{\bf v}_{2c\pm } &  \frac{\sqrt{3}}{2}(\hat\ax_{\rm d}\hat{\bf v}_\pm^t + \hat{\bf v}_\pm\hat\ax_{\rm d}^t)  
\\ 
\pm 3i &    \widetilde{\bf v}_{3\pm}   &  \frac1{\sqrt{2}}(\hat{\bf u}_\pm\hat{\bf v}_\pm^t + \hat{\bf v}_\pm\hat{\bf u}_\pm^t)  
\\ 
\pm 4i &  \widetilde{\bf v}_{4\pm} & \hat{\bf v}_\pm \hat{\bf v}_\pm^t
\end{matrix}
\eeq
Apart from the last two, which have unit multiplicity, these are not unique.  For instance, different combinations of the eigenvectors with multiplicity greater than one  could be used instead.    We also note from their definitions in eqs. \rf{ev3} and   \rf{ev6} that
\beq{090}
{\bf v}_{\pm}^* = - {\bf v}_{\mp}, 
\qquad
\hat{\bf u}_{\pm}^* = - \hat {\bf u}_{\mp}, 
\qquad
\hat{\bf v}_{\pm}^* =  \hat {\bf v}_{\mp} . 
\eeq
Hence, the five null vectors of $\widetilde{\bf P}$ in \rf{ev21} are real. 
$\widetilde{\bf P}$ can be expressed in terms of the remaining 16 eigenvectors as
\beq{081}
\widetilde{\bf P} =  \widetilde{\bf P}_1+ 2\widetilde{\bf P}_2+3\widetilde{\bf P}_3+4\widetilde{\bf P}_4, 
 \eeq
 where
 \beq{0819}
\widetilde{\bf P}_j =  \sum\limits_\pm\, \pm i   \sum\limits_{x =a,b,c} \widetilde{\bf v}_{jx\pm }\widetilde{\bf v}_{jx\pm }^*  , \ \ j=1,2;  
\qquad\qquad
\widetilde{\bf P}_j =  \widetilde{\bf v}_{j\pm }\widetilde{\bf v}_{j\pm }^*, \ \ j=3,4.
 \eeq
The identity is  
 \beq{081a}
\widetilde{\bf I} = \sum\limits_{x =a,b,c,d,e} \widetilde{\bf v}_{0x }\widetilde{\bf v}_{0x } +\sum\limits_\pm\,\bigr(   \sum\limits_{x =a,b,c} \big( \widetilde{\bf v}_{1x\pm }\widetilde{\bf v}_{1x\pm }^*  
+ \widetilde{\bf v}_{2x\pm }\widetilde{\bf v}_{2x\pm }^*  \big)
+  \widetilde{\bf v}_{3\pm }\widetilde{\bf v}_{3\pm }^*
+  \widetilde{\bf v}_{4\pm }\widetilde{\bf v}_{4\pm }^*
\bigr) \, . 
 \eeq
It is straightforward  to demonstrate using these expressions  that  $\widetilde{\bf P}$ satisfies 
 \beq{44}
  \widetilde{\bf P} \, ( \widetilde{\bf P}^2+1) \,  ( \widetilde{\bf P}^2+4) \,  ( \widetilde{\bf P}^2+9) \,( \widetilde{\bf P}^2+16)  = 0\, ,  
 \eeq
 which is the characteristic equation for $2n$-th order $P$ tensors, eq. \rf{13}.

\section{Projection onto TI symmetry and Cartan decomposition}\label{Discussion}

It has been mentioned in passing that  null vectors of $\tens P$ and its particular realizations for elasticity tensors correspond to  tensors with transversely isotropic symmetry.  These are defined as tensors invariant under the action of the group $SO(2)$ associated with rotation about the axis $\bd p$.
We conclude by making this connection more specific,   and relating the theory  for the rotation of tensors to the  Cartan decomposition, which is introduced below.

\subsection{Cartan decomposition} 

Referring to Theorem \ref{thm3}, the  expansion of the projector ${\tens M}_k$ defined in eq. \rf{640} 
  as an even  polynomial of degree $2n$ in ${\tens P}$, i.e. of  degree $n$ in ${\tens P}^2$, follows from  eqs. \rf{3450} and \rf{63} and the identity ${\tens P}_k^2 = k^{-1}{\tens P}{\tens P}_k$. 
Note that the  ${\tens M}_k$ tensors partition the identity
\beq{642} 
{\tens I} = {\tens M}_0+  {\tens M}_1+ \ldots +   {\tens M}_n.
\eeq
Starting with ${\tens M}_0$ we have 
\beq{204}
{\tens P}{\tens M}_0 = {\tens M}_0{\tens P}= 0, 
\eeq
indicating that ${\tens M}_0$ is the linear projection operator onto the null space of ${\tens P}$.
In order to examine the remaining $n$ projectors, note that   the rotation can be expressed
\beq{642a} 
{\tens Q}  = {\tens M}_0  + 
\sum\limits_{j=1}^n\,  \big( \cos  j\theta {\tens M}_j + \sin j\theta \,{\tens P}_j   \big) \, . 
\eeq
The action of $\tens Q$ on an $n$-th order tensor $\tens T$, the rotation of $\tens T$, can therefore be described as
\beq{643}
{\tens Q}{\tens T}  = {\tens T}_0  + 
\sum\limits_{j=1}^n\,  \big( \cos  j\theta {\tens T}_j  + \sin j\theta {\tens R}_j    \big) \, ,  
\eeq
where ${\tens T}_0 $ and ${\tens T}_j,  \, {\tens R}_j$, $j=1,2,\ldots, n$ are   defined by \rf{644}. 
We note that 
${\tens R}_j $  may be expressed 
\beq{050}
{\tens R}_j  = {\tens P}_j{\tens T}_j= \frac1{j} {\tens P}{\tens T}_j, \quad j=1,2,\ldots, n.
\eeq
Also,  ${\tens T}_0$ is  the component of ${\tens T}$ in the null space of ${\tens P}$, i.e. the transversely isotropic part of ${\tens T}$. 
Equation \rf{643} show that ${\tens T}_0 $ and the pairs ${\tens T}_j,{\tens R}_j$, j=1,2,\ldots, n, 
rotate separately, forming distinct subspaces, and hence proving Theorem \ref{thm3}.

\subsection{Application to elasticity tensors} 

The  projection operators are now illustrated with particular application to tensors of low order and elasticity tensors, starting with the simplest case: rotation of vectors.  

\subsubsection{Vectors, n=1}
If $n=1$, we have ${\tens T} = {\bf P}$ of eq. \rf{024}, and hence 
\beq{87}
{\bf M}_0 = \ax \ax , \qquad {\bf M}_1= {\bf I} - \ax \ax. 
\eeq
Also,  ${\tens T}\rightarrow {\bd t}$, an arbitrary vector, and the decomposition \rf{643} and \rf{644} is  
\beq{645}
{\bd Q}{\bd t} = ({\bd t}\cdot{\bd p}){\bd p} + \cos \theta\, 
[  {\bd t} - ({\bd t}\cdot{\bd p}){\bd p}] + 
\sin\theta\, \, {\bd p} {\bd \times}{\bd t} . 
\eeq
Thus, we identify ${\bd t}_0 = {\bf M}_0 {\bd t}$  as the component of the vector  parallel to the axis of rotation, ${\bd t}_1 = {\bf M}_1 {\bd t}$ as the  orthogonal complement, and ${\bd r}_1 = {\bf P}_1 {\bd t}$ is the rotation of ${\bd t}_1$ about the axis by $\pi/2$. 

\subsubsection{Second order tensors, n=2}
For $n=2$, we have ${\tens P}_1$, ${\tens P}_2$  given in \rf{00821} and 
\beq{821}
 {\tens M}_0 =({\tens P}^2+1)({\tens P}^2+4)/4, 
 \qquad
  {\tens M}_1 = -{\tens P}^2 ({\tens P}^2+4)/3, 
 \qquad
  {\tens M}_2= {\tens P}^2 ({\tens P}^2+1)/12.  
\eeq
The dimensions of the subspaces are ${\tens T}_0, \, 3$;  
${\tens T}_1,\, 2$, ;  ${\tens R}_1,\, 2$; 
${\tens T}_2,\, 1$, ;  ${\tens R}_2 ,\, 1$.

For   {\emph symmetric} second order tensors   the projectors $\whbf{M}_0$, $\whbf{M}_1$ and 
$\whbf{M}_2$ have the same form as in \rf{821} in terms of $\whbf{P}$, and may also be expressed in terms of the fundamental vectors introduced earlier.  Thus, using 
\rf{p13} and \rf{p133}, 
\beq{330}
\whbf{M}_0 =  \frac14(\whbf{P}^2 +\whbf{I})(\whbf{P}^2 +4\whbf{I})
= 
\hat{\bf i}\hat{\bf i}^t + \hat\ax_{\rm d}\hat\ax_{\rm d}^t \, , 
\qquad
\whbf{M}_1 = 
\hat{\bf u}_+ \hat{\bf u}_+^*  +  \hat{\bf u}_-\hat{\bf u}_-^*, 
\qquad
\whbf{M}_2 =
\hat{\bf v}_+ \hat{\bf v}_+^* +  \hat{\bf v}_-\hat{\bf v}_-^*. 
 \eeq
Using these and $\whbf{P}_1$, $\whbf{P}_2$ from \rf{p13}, it follows that the subspace associated with   
$\whbf{M}_0$ is two dimensional, while $\whbf{M}_1^{(1)}$, $\whbf{M}_1^{(2)}$, $\whbf{M}_2^{(1)}$, and $\whbf{M}_2^{(2)}$ have dimension one.  This Cartan decomposition of the $6$-dimensional space $Sym$ agrees with a different approach by Forte and   Vianello \cite{FV}.  Alternatively, 
using \rf{330} and the dyadics in \rf{ev6}, it follows that 
\begin{subequations}\label{2841}
\bal{2841a}
{\tens M}_0 &= \ax \ax \ax \ax +\frac12({\bf I} - \ax \ax)({\bf I} - \ax \ax), 
\\
{\tens M}_1 &= \frac12( {\bd p}{\bd q}+ {\bd q}{\bd p})( {\bd p}{\bd q}+ {\bd q}{\bd p})
+\frac12( {\bd p}{\bd r}+ {\bd r}{\bd p})( {\bd p}{\bd r}+ {\bd r}{\bd p}),
\\
{\tens M}_2 & = \frac12( {\bd q}{\bd q}- {\bd r}{\bd r})( {\bd q}{\bd q}- {\bd r}{\bd r})
+\frac12( {\bd q}{\bd r}+ {\bd r}{\bd q})( {\bd q}{\bd r}+ {\bd r}{\bd q}) . 
\end{align}
\end{subequations}

The action of ${\tens M}_0$ on a second order symmetric tensor is 
\beq{285}
{\tens M}_0{ \bd T}
= { \bd T}_0
=  T_\parallel  \ax \ax + T_\perp({\bf I} - \ax \ax), 
\eeq
where $T_\parallel = { \bd T}:{\bd p}{\bd p}$ and 
$T_\perp  = \frac12 ( \tr { \bd T} - { \bd T}:{\bd p}{\bd p})$. It is clear that ${ \bd T}_0$ is unchanged under   rotation about $\ax$.  Similarly, ${\tens M}_1{ \bd T}$ transforms as a vector under rotation, hence the suffix $1$, while  ${\tens M}_2{ \bd T}$ transforms as a second order tensor with elements confined to the plane orthogonal to the axis of rotation. 

Let  ${\bd p}= {\bd e}_3$, then using  $\rf{330}_1$ or the simpler \rf{2841a} yields 
\beq{332}
\hat{ \bd T}_0= 
\whbf{M}_0 ({\bd e}_3) \hat{ \bd T} 
=
\begin{pmatrix}
\frac12 & 0 & 0 & 0 & 0 & 0 \\ 
0 &\frac12  & 0 & 0 & 0 & 0 \\ 
0 & 0   & 1 & 0 & 0 & 0 \\ 
0 & 0   & 0 & 0 & 0 & 0 \\ 
0 & 0   & 0 & 0 & 0 & 0 \\ 
0 & 0   & 0 & 0 & 0 & 0 \end{pmatrix}\hat{ \bd T}  
=  \begin{pmatrix}
 (\hat{T}_{1}+\hat{T}_{2})/2
\\   
 (\hat{T}_{1}+\hat{T}_{2})/2
\\  
\hat{T}_{3}
\\ 0
\\  0
\\  0
\end{pmatrix} .
\eeq
The projection $\hat{ \bd T}_0$ can be identified as the part of $\hat{ \bd T}$ which displays hexagonal (or transversely isotropic) symmetry about the axis ${\bd e}_3$.  To be more precise, if we denote the projection $\whbf{M}_0  \hat{ \bd T} $ as $\hat{ \bd T}_{\rm Hex} $, 
then $\hat{ \bd T}_{\rm Hex} $ is invariant under the action of the symmetry group of rotations about ${\bd p}$.  It may also be shown that of all tensors with hexagonal symmetry $\hat{ \bd T}_{\rm Hex} $ is the closest to $\hat{ \bd T}$ using a Euclidean norm for distance.  Similarly, 
\beq{3322}
\hat{ \bd T}_1= 
\whbf{M}_1   \hat{ \bd T} 
= \big( 0,\,0,\,0,\, \hat{T}_4,\, \hat{T}_5,\, 0\big)^t,
\qquad
\hat{ \bd T}_2= 
\whbf{M}_2   \hat{ \bd T} 
= \big( \frac12(\hat{T}_{1}-\hat{T}_{2}) ,\,\frac12(\hat{T}_{2}-\hat{T}_{1}) ,\,0,\, 0,\, 0,\, \hat{T}_6\big)^t ,
\eeq
and using eqs. \rf{0250} and \rf{3322}, 
\beq{322}
\hat{ \bd R}_1 
= \big( 0,\,0,\,0,\, \hat{T}_5,\, -\hat{T}_4,\, 0\big)^t,
\qquad
\hat{ \bd R}_2 
= \big( -\frac1{\sqrt{2}}\hat{T}_6 ,\, \frac1{\sqrt{2}}\hat{T}_6 ,\,  0,\, 0,\, 0,\, 
\frac1{\sqrt{2}}(\hat{T}_{1}-\hat{T}_{2})
\big)^t .
\eeq
It can be checked that under rotation, we have 
\beq{089}
\hat{ \bd Q}\hat{ \bd T}_0 =  \hat{ \bd T}_0,
\qquad
\hat{ \bd Q}\hat{ \bd T}_j = \cos j\theta \hat{ \bd T}_j + \sin j\theta \hat{ \bd R}_j,
\qquad
\hat{ \bd Q}\hat{ \bd R}_j = \cos j\theta \hat{ \bd R}_j - \sin j\theta \hat{ \bd T}_j,
\quad j=1,2. 
 \eeq
 Thus, the singleton $\hat{ \bd T}_0$ and pairs $(\hat{ \bd T}_1, \hat{ \bd R}_1) $ and $(\hat{ \bd T}_2, \hat{ \bd R}_2) $ form closed groups under the action of the rotation.

\subsubsection{Elastic moduli, n=4}

The $5$-dimensional null space of $\widetilde{\bf P}$ is equal to the set of base elasticity  tensors for transverse isotropy 
\cite{Zheng93,FV,Lu98}.  
The projector follows from eqs. 
 \rf{0819} and \rf{081a} as  
\beq{331}
\widetilde{\bf M}_0  \equiv \frac{1}{(4!)^2}
( \widetilde{\bf P}^2+1) \,  ( \widetilde{\bf P}^2+4) \,  ( \widetilde{\bf P}^2+9) \,( \widetilde{\bf P}^2+16)
= \sum\limits_{x =a,b,c,d,e} \widetilde{\bf v}_{0x }\widetilde{\bf v}_{0x }  \, . 
 \eeq
This is equivalent to the 8-th order projection tensor $P_{\rm Hex}$ of eq. (68c) of Moakher and Norris
\cite{moakher06b}, who use the base tensors suggested by Walpole \cite{Walpole86}.  The latter representation is useful because Walpole's tensors have a nice algebraic structure, making tensor products simple. 
Alternatively, in the spirit of the $21\times 21$ representation of elastic moduli, $\rf{331}_1$ and \rf{213} yield 
\beq{333}
\widetilde{M}_0 ({\bd e}_3) = \begin{pmatrix}
M_{\rm Hex} & {\bf 0}_{9\times 12}
\\ \\ 
 {\bf 0}_{12\times 9}& {\bf 0}_{12\times 12} 
 \end{pmatrix}, 
 \qquad 
M_{\rm Hex} = \begin{pmatrix}
\frac38~ & ~\frac38~ & ~0~ & ~0~ & ~0~ & ~\frac1{4\sqrt{2}}~& ~0~ & ~0~ & ~\frac14 
\\ 
\frac38 & \frac38 & 0 & 0 & 0 & \frac1{4\sqrt{2}}& 0 & 0 & \frac14 
\\ 
0 & 0 & 1 &  0 & 0 & 0 &  0 & 0 & 0 
\\ 
0 & 0 & 0 & \frac12 &   \frac12 & 0 &  0 & 0 & 0   
\\ 
0 & 0 & 0 & \frac12 &   \frac12 & 0 &  0 & 0 & 0 
\\ 
\frac1{4\sqrt{2}}& \frac1{4\sqrt{2}}& 0 &  0 & 0 & \frac34 & 0 &  0 & -\frac1{2\sqrt{2}}
\\ 
0 & 0 &  0 & 0 & 0 & 0 & \frac12 & \frac12 &0
\\ 
0 & 0 &  0 & 0 & 0 & 0 & \frac12 & \frac12 &0
\\ 
\frac14  & \frac14 & 0 & 0 & 0 & -\frac1{2\sqrt{2}} &0 & 0 &\frac12
\end{pmatrix}\, .
\eeq
The projection $\widetilde{M}_0$ of \rf{333} 
 is identical to the   matrix  derived by Browaeys and Chevrot \cite{Browaeys04} for projection of elastic moduli onto hexagonal symmetry.  Specifically, the $9\times 9$ matrix $M_{\rm Hex}$ is exactly ${\bd M}^{(4)}$ of Browaeys and Chevrot (see Appendix A of \cite{Browaeys04}).  
Hence, $\widetilde{M}_0$ of \rf{333} provides an algorithm for projection onto hexagonal symmetry for arbitrary axis ${\bd p}$. 

The subspace  of $\widetilde{T}_0$ is $5$-dimensional, while the subspaces $( \widetilde{T}_1, \widetilde{R}_1)$, $( \widetilde{T}_2, \widetilde{R}_2)$, $( \widetilde{T}_3, \widetilde{R}_3)$  and $( \widetilde{T}_4, \widetilde{R}_4)$ have dimensions $6$, $6$, $2$ and $2$, respectively.  These may be evaluated for arbitrary axis of rotation $\ax$ using the prescription in eqs. \rf{641c} and \rf{050}.




\renewcommand{\appendix}{\setcounter{section}{0}\renewcommand{\thesection}{}
\renewcommand{\thesection}{\Alph{section}.}
\setcounter{equation}{0}
 			\renewcommand{\theequation}{\thesection\arabic{equation}}
\section{Appendix: A recursion }
}
  			
\appendix

Equation \rf{05030} provides an expansion in terms of  basic trigonometric functions. An alternative recursive procedure is  derived here.  Consider the expression  \rf{05030} written  
\beq{512}
e^{B} = f_m (B)\, ,  
\eeq
where the function $f_m (B)$ is a finite series defined by the $m$ distinct pairs of non-zero eigenvalues of $B$. 
Suppose the set of  eigenvalues is augmented by one more pair, $\pm i c_{m+1}$, with the others unchanged.  Then the sums and products in \rf{05030} change to reflect the new eigenvalues, 
\beq{513}
e^{B} = f_{m+1} (B) 
 = I  + 
\sum\limits_{j=1}^{m+1}\,  
\big[ c_j \sin   c_j  \, B + ( 1- \cos   c_j ) \,B^2  \big] 
\, c_j^{-2} \, 
  \prod\limits_{\substack{k=1 \\ k\ne j}}^{m+1} \bigr( 
  \frac{B^2 + c_k^2 } {c_k^2 - c_j^2 }\bigr) \, .
\eeq
The effect of the two additional eigenvalues can be isolated using the identity 
\beq{514}
 \frac{B^2 + c_{m+1}^2}{c_{m+1}^2 - c_j^2} = 1 + \frac{B^2 + c_j^2}{c_{m+1}^2 - c_j^2}, 
\eeq
to give 
\beq{515}
 f_{m+1} = f_m 
 + \bigg( 
\sum\limits_{j=1}^{m+1}\,  
\frac{  c_j \sin   c_j  \, B + ( 1- \cos   c_j ) \, B^2  }
{ c_j^{2} \prod\limits_{\substack{l=1 \\ l\ne j}}^{m+1} ( c_l^2 - c_j^2) }
\bigg)
 \,  \prod\limits_{k=1}^{m} (B^2 + c_k^2) \, .
\eeq
This defines the  sequence, starting with $f_0 = I$. 

Applying this to the rotation $\exp ( \theta {\tens P})$, and noting that the eigenvalues of the skew symmetric tensor $\tens P$ correspond to 
$c_j = j\theta$, gives  
\beq{517}
P_{2n+2}(\theta, x) = P_{2n}(\theta, x)
 + \bigg( 
\sum\limits_{j=1}^{n+1}\,  
\frac{  j \sin j\theta    \, x + ( 1- \cos j\theta ) \, x^2  }
{ j^{2} \prod\limits_{\substack{l=1 \\ l\ne j}}^{n+1} ( l^2 - j^2) }
\bigg)
 \,   \prod\limits_{k=1}^{n} (x^2 + k^2) \, .
\eeq
The sum over trigonometric functions may be simplified using formulas for $\sin n\theta$ and $\cos n\theta$ in terms of  $\cos \theta/2$ powers of  $\sin \theta/2$ in Section 1.33 of Gradshteyn et al. \cite{Gradshteyn}. 
Thus, 
\bal{518}
\sum\limits_{j=1}^{n+1}\,  
\frac{   \sin j\theta     }
{ j \prod\limits_{\substack{l=1 \\ l\ne j}}^{n+1} ( l^2 - j^2) }
&=\frac{ (2\sin\frac{\theta}{2})^{2n+1} }{(2n+1)!}\, \cos\frac{\theta}{2} \,  ,
\\
\sum\limits_{j=1}^{n+1}\,  
\frac{  1- \cos j\theta  }
{ j^{2} \prod\limits_{\substack{l=1 \\ l\ne j}}^{n+1} ( l^2 - j^2) }
&=  \frac{ (2\sin\frac{\theta}{2})^{2n+1} }{(2n+1)!}\, \frac{\sin\frac{\theta}{2}}{n+1}\, ,
\end{align}
and hence, 
\beq{519}
P_{2n+2}(\theta, x) = P_{2n}(\theta, x)
 + \frac{ (2\sin\frac{\theta}{2})^{2n+1} }{(2n+1)!}\,
 \big( \cos\frac{\theta}{2} \, x + (n+1)^{-1}\sin\frac{\theta}{2} x^2)
 \,   \prod\limits_{k=1}^{n} (x^2 + k^2) \, ,
\eeq
from which the expression \rf{5191b} follows.



\newpage   \setlength{\topmargin}{-0.6in}
\rotatebox{90}{ 
\begin{minipage}[b]{9.in} 
{\tiny 
\begin{align*}
\left( \begin {array}{ccccccccccccccccccccc} 0&0&0&0&0&0&0&0&0&0&0&0&0&2\,p_2&0&0&0&-2\,p_3&0&0&0\\\noalign{\medskip}0&0&0&0&0&0&0&0&0&0&0&0&0&0&2\,p_3&-2\,p_1&0&0&0&0&0\\\noalign{\medskip}0&0&0&0&0&0&0&0&0&0&0&0&2\,p_1&0&0&0&-2\,p_2&0&0&0&0\\\noalign{\medskip}0&0&0&0&0&0&0&0&0&0&-\sqrt {2}p_2&\sqrt {2}p_3&-\sqrt {2}p_1&0&0&\sqrt {2}p_1&0&0&0&0&0\\\noalign{\medskip}0&0&0&0&0&0&0&0&0&\sqrt {2}p_1&0&-\sqrt {2}p_3&0&-\sqrt {2}p_2&0&0&\sqrt {2}p_2&0&0&0&0\\\noalign{\medskip}0&0&0&0&0&0&0&0&0&-\sqrt {2}p_1&\sqrt {2}p_2&0&0&0&-\sqrt {2}p_3&0&0&\sqrt {2}p_3&0&0&0\\\noalign{\medskip}0&0&0&0&0&0&0&0&0&0&0&0&-2\,p_1&0&0&2\,p_1&0&0&0&-\sqrt {2}p_2&\sqrt {2}p_3\\\noalign{\medskip}0&0&0&0&0&0&0&0&0&0&0&0&0&-2\,p_2&0&0&2\,p_2&0&\sqrt {2}p_1&0&-\sqrt {2}p_3\\\noalign{\medskip}0&0&0&0&0&0&0&0&0&0&0&0&0&0&-2\,p_3&0&0&2\,p_3&-\sqrt {2}p_1&\sqrt {2}p_2&0\\\noalign{\medskip}0&0&0&0&-\sqrt {2}p_1&\sqrt {2}p_1&0&0&0&0&0&0&0&p_3&0&0&0&-p_2&0&-\sqrt {2}p_3&\sqrt {2}p_2\\\noalign{\medskip}0&0&0&\sqrt {2}p_2&0&-\sqrt {2}p_2&0&0&0&0&0&0&0&0&p_1&-p_3&0&0&\sqrt {2}p_3&0&-\sqrt {2}p_1\\\noalign{\medskip}0&0&0&-\sqrt {2}p_3&\sqrt {2}p_3&0&0&0&0&0&0&0&p_2&0&0&0&-p_1&0&-\sqrt {2}p_2&\sqrt {2}p_1&0\\\noalign{\medskip}0&0&-2\,p_1&\sqrt {2}p_1&0&0&2\,p_1&0&0&0&0&-p_2&0&0&0&0&p_3&0&0&0&-\sqrt {2}p_2\\\noalign{\medskip}-2\,p_2&0&0&0&\sqrt {2}p_2&0&0&2\,p_2&0&-p_3&0&0&0&0&0&0&0&p_1&-\sqrt {2}p_3&0&0\\\noalign{\medskip}0&-2\,p_3&0&0&0&\sqrt {2}p_3&0&0&2\,p_3&0&-p_1&0&0&0&0&p_2&0&0&0&-\sqrt {2}p_1&0\\\noalign{\medskip}0&2\,p_1&0&-\sqrt {2}p_1&0&0&-2\,p_1&0&0&0&p_3&0&0&0&-p_2&0&0&0&0&\sqrt {2}p_3&0\\\noalign{\medskip}0&0&2\,p_2&0&-\sqrt {2}p_2&0&0&-2\,p_2&0&0&0&p_1&-p_3&0&0&0&0&0&0&0&\sqrt {2}p_1\\\noalign{\medskip}2\,p_3&0&0&0&0&-\sqrt {2}p_3&0&0&-2\,p_3&p_2&0&0&0&-p_1&0&0&0&0&\sqrt {2}p_2&0&0\\\noalign{\medskip}0&0&0&0&0&0&0&-\sqrt {2}p_1&\sqrt {2}p_1&0&-\sqrt {2}p_3&\sqrt {2}p_2&0&\sqrt {2}p_3&0&0&0&-\sqrt {2}p_2&0&-p_3&p_2\\\noalign{\medskip}0&0&0&0&0&0&\sqrt {2}p_2&0&-\sqrt {2}p_2&\sqrt {2}p_3&0&-\sqrt {2}p_1&0&0&\sqrt {2}p_1&-\sqrt {2}p_3&0&0&p_3&0&-p_1\\\noalign{\medskip}0&0&0&0&0&0&-\sqrt {2}p_3&\sqrt {2}p_3&0&-\sqrt {2}p_2&\sqrt {2}p_1&0&\sqrt {2}p_2&0&0&0&-\sqrt {2}p_1&0&-p_2&p_1&0
\end {array} \right) 
\end{align*} 
} 
\bigskip 
\begin{center}
Table 1. The matrix $\widetilde{\bf P}( \ax)\in so(21)$ for $4$th order elasticity tensors represented as $21$-vectors.  The rotation $\widetilde{\bf Q} = \exp (\theta \widetilde{\bf P})$ is given by the 8-th order polynomial $\widetilde{\bf Q} = P_8( \theta , \widetilde{\bf P})$ of eq. \rf{433}. 
\end{center}
  \end{minipage}
}
\newpage
\end{document}